\documentclass[12pt]{article}
\usepackage[margin=1.0in]{geometry}
\usepackage[utf8]{inputenc}
\usepackage[intlimits]{amsmath}
\usepackage{amssymb,amsfonts,amsthm,array,dsfont}
\usepackage{physics}
\usepackage[english]{babel}
\usepackage{rotating}
\usepackage{mathrsfs}
\usepackage{empheq}
\usepackage{breqn}
\usepackage[bottom]{footmisc}

\usepackage{csquotes}

\renewcommand{\d}{\mbox{d}}
\renewcommand{\bar}[1]{\overline{#1}}

\renewcommand{\dd}{\partial}
\newcommand{\rhs}{{\it r.h.s.} }

\numberwithin{equation}{section}

\begin{document}

	\vspace{1.7 cm}
	
	\begin{flushright}
		
		{\small FIAN/TD/15-2022}
	\end{flushright}
	\vspace{1.7 cm}
	
	\begin{center}
		{\large\bf Shifted Homotopy Analysis of the Linearized Higher-Spin Equations in Arbitrary Higher-Spin Background}
		
		\vspace{1 cm}
		
		{\bf A.A.~Tarusov$^{1,2}$, K.A.~Ushakov$^{1,2}$ and  M.A.~Vasiliev$^{1,2}$}\\
		\vspace{0.5 cm}
		\textbf{}\textbf{}\\
		\vspace{0.5cm}
		\textit{${}^1$ I.E. Tamm Department of Theoretical Physics,
			Lebedev Physical Institute,}\\
		\textit{ Leninsky prospect 53, 119991, Moscow, Russia}\\
		
		\vspace{0.7 cm} \textit{
			${}^2$ Moscow Institute of Physics and Technology,\\
			Institutsky lane 9, 141700, Dolgoprudny, Moscow region, Russia
		}

            \vspace{0.7 cm} sasha.tarusov@gmail.com\,, ushakovkirill@mail.ru\,, vasiliev@lpi.ru
		
	\end{center}
	
	\vspace{0.4 cm}
	
	\begin{abstract}
		\noindent
		Analysis of the first-order corrections to higher-spin equations is extended to homotopy operators involving shift parameters with respect to the spinor $Y$ variables, the argument of the higher-spin connection $\omega(Y)$ and the argument of the higher-spin zero-form $C(Y)$. It is shown that a relaxed uniform $(y+p)$-shift and a shift by the argument of $\omega(Y)$ respect the proper form of the free higher-spin equations and 
constitute a one-parametric class of vertices that contains those resulting from the conventional (no shift) homotopy. A pure shift by the argument of $\omega(Y)$ is shown not to affect the one-form  higher-spin field $W$ in the first order and, hence, the form of the respective vertices.

	\end{abstract}
	
	\newpage
	
	\vspace{-1cm}
	\tableofcontents
	
	\newpage
	
	\section{Introduction}
	
	Higher-spin (HS) gauge theory describes an infinite tower of gauge fields of all spins. Nonlinear field equations for $4d$ massless fields of all spins were found in \cite{Vasiliev:1990en,Vasiliev:1992av}. They admit $AdS_4$ as the most symmetric vacuum solution. The presence of $AdS_4$ radius as a dimensionful parameter in HS vertices potentially allows an infinite number of higher-derivative terms. Because of this HS gauge theory is not a local field theory in the usual sense. Instead of space-time locality, spin-locality (that is locality for any finite subset of fields) in the space of auxiliary
	spinor variables can be achieved at least in the lowest orders  \cite{Vasiliev:2015wma,Gelfond:2018vmi,
Didenko:2018fgx, Didenko:2019xzz, Gelfond:2019tac, Didenko:2020bxd, Gelfond:2021two}. In the lowest order, spin-locality in the spinor space is equivalent to space-time spin-locality. The conditions allowing to extend
 this property to higher orders were found recently in \cite{Vasiliev:2022med}.
	
	In the approach of  \cite{Vasiliev:1989}, HS fields in $AdS_4$ are described by the one-form
	$\omega(Y;K|x)$ and zero-form $C(Y;K|x)$ that depend on space-time coordinates $x$,
auxiliary variables $Y_A = (y_{\mu}, \bar{y}_{\dot{\mu}})$,
$\mu, \dot{\mu} = 1,2$, and Klein operators $K$. Both $\omega(Y;K|x)$ and $C(Y;K|x)$ are
regular functions of $Y^A$ that serve
as the generating functions for the component fields
\begin{equation}
F(Y; K|x) = \sum_{n=0}^\infty \sum_{m=0}^\infty \frac{1}{n!m!} F^{\mu_1 ... \mu_n,}{}^{\dot\mu_1 ... \dot\mu_m} (K|x) y_{\mu_1} ... y_{\mu_n} \bar y_{\dot\mu_1} ... \bar y_{\dot\mu_m} \,,
\end{equation}
$F=\omega(Y;K|x)$ or $C(Y;K|x)$.
The  Klein operators $K$ induce the field doubling that does not matter in the
consideration of this section (for more
detail see \cite{Vasiliev:1989,Vasiliev:1999ba} and Section \ref{HSE}).
	
	Unfolded form of the free HS equations in the gauge sector
   referred to	as First On-Shell Theorem is \cite{Vasiliev:1989}
	(for detailed recent analysis see \cite{Bychkov:2021zvd})
	\begin{equation} \label{e: COST}
		R^{\mu(n),\dot\mu(m)}(x)=\delta_{0,n}\,h_{\nu\dot\mu} h^{\nu}_{\ \dot\mu}\bar C^{\dot\mu(m+2)}(x) + \delta_{0, m}\, h_{\mu\dot\nu} h_{\mu}^{\ \dot\nu} C^{\mu(n+2)}(x)\,,
	\end{equation}
	where only exterior products of differential forms are used (from now on the wedge symbol is implicit)
and\footnote{We  use a shorthand notation $\omega_{\mu(n),\dot\mu(m)}(x) = \omega_{\mu_1...\mu_n, \dot \mu_1 ... \dot \mu_m}(x)$ for totally
symmetric multispinors. Spinor indices are raised and lowered according to the rules $A^\mu=\epsilon^{\mu\nu}A_\nu$,
$A_\mu=A^\nu\epsilon_{\nu\mu}$, $\epsilon_{\nu\mu}=-\epsilon_{\mu\nu}$, $\epsilon_{12}=\epsilon^{12}=1$ and analogously for
dotted indices.}
	\begin{equation}
		R^{\mu(n),\dot\mu(m)}(x):=D_L\omega^{\mu(n),\dot\mu(m)}(x) + \lambda(n h^\mu_{\ \dot\rho}(x) \omega^{\mu(n-1),\dot\rho\dot\mu(m)}(x) + m h_\rho^{\ \dot\mu} (x) \omega^{\rho\mu(n),\dot\mu(m-1)} (x))\,,
	\end{equation}
	where $\lambda$ is the inverse $AdS$ radius, and $D_L=\d_x+\varpi +\bar\varpi$ is a Lorentz-covariant derivative, with   space-time de Rham derivative $\d_x$ and
	Cartan's spin-connection $(\varpi\oplus\bar\varpi)$ and
	\begin{equation*}
		D_L \omega_{\mu(n),\dot\mu(m)}(x):=\d_x \omega_{\mu(n),\dot\mu(m)}(x)+n\varpi_\mu^{\ \nu}(x)\omega_{\nu\mu(n-1),\dot\mu(m)}(x) + m\bar\varpi_{\dot\mu}^{\ \dot\nu}(x)\omega_{\mu(n),\dot\nu\dot\mu(m-1)}(x)\,.
	\end{equation*}

HS equations are formulated in terms of  the zero-forms $C(Y;K|x)$ and one-forms $\omega(Y;K|x)$.
 The field variables associated with spin $s = 0, 1/2, 1, 3/2, ...$ are
	\begin{equation}\label{e: components_meaning}
		\omega^{\mu(n), \dot \mu(m)}(x) : n + m = 2(s-1)\,,\qquad C^{\mu(n), \dot \mu(m)} (x): |n - m| = 2s\,.
	\end{equation}
	Fronsdal  fields are described in terms of the generalized frame one-form
$\omega^{\mu(n), \dot \mu(m)}(x)$  with $n = m$ for bosons and $\abs{n - m} = 1$ for fermions,
the scalar $C(x)$, and the pair of spin $1/2$ fields: $C^\mu(x)$, $C^{\dot{\mu}}(x)$. Fields with other values of $n,m$  describe derivatives of the Fronsdal field. Specifically, zero-forms $C(Y;K|x)$ describe gauge invariant combinations of derivatives of the   Fronsdal fields (linearized curvatures) resulting
  from equation (\ref{e: COST}) and the equation
	\begin{equation}
		\tilde{D} C(Y;K|x) := \bigg(D_L - \lambda h^{\mu \dot \mu}\big(y_\mu \bar y_{\dot \mu} + \frac{\partial^2}{\partial y^\mu \partial \bar y^{\dot \mu}} \big)\bigg)C(Y;K|x) = 0\,, \label{DC}
  \end{equation}
 where
 \begin{equation}
 C(Y;K|x) = - C(Y;-K|x)\,,
\end{equation}
which along with (\ref{e: COST}) form a full set of free HS equations for all massless
fields in $AdS_4$.

	For any fixed spin $s$, the maximal number of derivatives of the Fronsdal field contained in
$\omega^{\mu(n), \dot \mu(m)}(x)$
and  $C^{\mu(n), \dot \mu(m)}(x)$ is $[s]-1$ and
$\frac{n+m}{2} - \{s\}$, respectively.
 Along with (\ref{e: components_meaning}) this implies that for each  spin $s$ there is a
finite number of fields in $\omega(Y;K|x)$ and an infinite number of fields in $C(Y;K|x)$,
which is the  source of potential non-locality.
	
	For the analysis of locality it is important to preserve the form (\ref{e: COST}) of the free HS equations.
	Indeed, not every scheme of  perturbative analysis of the nonlinear HS equations automatically
reproduces free equations	in the form (\ref{e: COST}).
As discussed in \cite{Vasiliev:1989}, it may deform the \rhs of (\ref{e: COST}) bringing to it
other components of $C(Y;K|x)$. In such a case to reproduce
	the First On-Shell Theorem it is necessary to make a field redefinition of the zero-forms $C(Y;K|x)$,
	the physical meaning of which  is obscure from the locality perspective. As a consequence,
	the analysis of (non-)locality of the higher-order vertices is obstructed either until a field redefinition
	bringing free HS equations to the form of First On-Shell Theorem is performed. Thus, it is
	vital to keep linearized HS equations  in the form (\ref{e: COST}) applying only such field
	redefinitions that respect the First On-Shell Theorem.
	
	A useful way to analyse HS equations  is to reconstruct interacting
	vertices in the unfolded form \cite{Vasiliev:1989}
	\begin{equation}
		\d_x \omega = - \omega * \omega + \Upsilon(\omega, \omega, C) + \Upsilon(\omega, \omega, C, C) + ...\,, \label{e: verteq_1}
	\end{equation}
	\begin{equation}
		\d_x C = - [\omega, C ]_* + \Upsilon(\omega, C, C) + ...\,, \label{e: verteq_2}
	\end{equation}
	where
$*$ denotes the Moyal star product underlying the HS algebra \cite{Vasiliev:1986qx}
\begin{equation}
		f(Y)*g(Y) = f(Y) e^{i \epsilon^{A B} \overleftarrow{\partial}_A \overrightarrow{\partial}_B} g(Y) \,. \label{y_star_product}
	\end{equation}
(See \cite{Vasiliev:1999ba} for a review and more references.)

In the formulation of HS equations of \cite{Vasiliev:1992av}
the derivation of the interaction vertices amounts to solving first-order differential
equations with a nilpotent differential in the auxiliary spinor space.  At each order one faces the
 cohomological freedom that effectively encodes the field  redefinitions. The choice of
one or another cohomology class is determined by the choice of the homotopy operator that resolves the differential equation in the spinor space. Though properly reproducing free HS equations, the
seemingly most natural  conventional homotopy of \cite{Vasiliev:1992av}
	leads to non-local vertices starting from the second order \cite{Boulanger:2015ova}.
	In \cite{Gelfond:2018vmi} it was suggested that the proper approach is based on the shifted homotopy,
 allowing to decrease the level of non-locality at higher orders, the technique further developed in \cite{Didenko:2018fgx}.
	The shifted homotopy operators involve the shifts of arguments of the dynamical HS fields $\omega(Y;K|x)$ and $C(Y;K|x)$ with some parameters.
	 The shifted homotopy technique was proven to be efficient by simplifying the analysis of locality
 of lower-order vertices \cite{Didenko:2018fgx, Didenko:2019xzz, Didenko:2020bxd,Gelfond:2021two} (for detail see Section \ref{PERT}). However, in these papers only the class of shifted homotopy operators dependent on the derivatives with respect to the spinor
 arguments $Y$  of  $C(Y;K|x)$ was considered.

 The goal of this paper is to fill in this gap by considering a more general class  of shifted homotopy operators involving the  derivatives with respect to the spinor arguments  of the HS one-forms $\omega(Y;K|x)$. An important condition of the linearized  analysis in the zero-forms $C(Y;K|x)$ (which parameterize linearized gauge invariant HS curvatures) considered in this paper is that it should not affect the form  of free HS equations (\ref{e: COST}) since, otherwise, this would spoil the interpretation of the zero-forms $C(Y;K|x)$  in terms of derivatives of the HS gauge fields $\omega(Y;K|x)$ ruining the higher-order locality analysis in terms of  $C(Y;K|x)$. The particular form of the \rhs of (\ref{e: COST}) is closely tied to the proper choice of field variables.
	In this paper we study the impact of  $\omega$-shifts by the argument of  $\omega(Y;K|x)$  on the
 vertex	$\Upsilon(\omega, \omega, C)$ within the shifted homotopy approach and test a more general linear shift in the homotopy procedures compared to the ones currently available in the literature, for instance in \cite{Didenko:2018fgx}.
 Such  shifts do not affect locality of the vertex unless they
	change the form of the \rhs of (\ref{e: COST}), which is undesirable as explained above.
	Our goal is to find the class of shift parameters that preserve the form of the
	 First On-Shell Theorem. Surprisingly, we find that
	the free $\omega$-shift parameters not only respect the form of (\ref{e: COST}) but, in the case of pure $\omega$-shifts, also
do not affect the perturbative corrections to the HS fields, and thus  the nonlinear vertices either, being equivalent  to those resulting from the homotopy with zero $\omega$-shift parameters.We also find that shifts with respect to $Y$ variables or  derivatives $p$ of the $Y$ arguments of $C(Y;K|x)$ can be present in a relaxed uniform way, that is their shift parameters must be equal to each other at each homotopy procedure step, but not necessarily shared between the procedures of different homotopy steps as in \cite{Didenko:2018fgx}. Such shifts preserve the form of the First On-Shell Theorem in the $AdS_4$ background but produce a one-parametric class of pairwise different vertices  for general background HS gauge one-forms.

The paper is organized as follows: in Section \ref{HSE}, the structure of the HS equations is
briefly recalled, perturbative analysis of which is recalled in Section \ref{PERT}.
Section \ref{Shifted homotopies} summarizes
 the key properties of the shifted homotopy technique relevant to our analysis. These results are
 then applied in Section \ref{Mod Shift} to the derivation of the form of the vertices with the
 shifts acting on the argument of $\omega(Y;K|x)$. The possible values of the  $y$- and $p$-shift  parameters
  that preserve the form of the First On-Shell Theorem are deduced in Section \ref{Adm Param}, while the effect of  the pure $\omega$-shift is investigated in Section \ref{Omega Shift}. Section \ref{Concl} contains a brief conclusion.

	\section{Higher-spin equations}\label{HSE}
	
	In the frame-like approach to HS equations in $AdS_4$, dynamics of the system is encoded in a one-form $\omega(y, \bar y; K | x)$ and a zero-form $C(y, \bar y; K | x)$ which are regular functions  of $sp(4)$ spinors $Y_A = (y_{\mu}, \bar{y}_{\dot{\mu}})$. $K=(k,\bar k)$ is a pair of Klein operators which will be introduced shortly.
	HS algebra is defined via  Moyal  star product (\ref{y_star_product})
with the  $sp(4)$-invariant form $\epsilon^{AB}=(\epsilon^{\mu \nu}, \epsilon^{\dot{\mu} \dot{\nu}})$,
	generated  by the relations
	\begin{equation}
		[y_\mu, y_\nu]_* = 2i \epsilon_{\mu \nu}\,, \qquad [\bar{y}_{\dot{\mu}}, \bar{y}_{\dot{\nu}}]_* = 2i \epsilon_{\dot{\mu} \dot{\nu}}\,, \qquad [y_\mu, \bar{y}_{\dot{\nu}}]_* = 0\,.
	\end{equation}

	Following \cite{Vasiliev:1992av} we introduce  auxiliary variables $Z_A=(z_\mu, \bar{z}
	_{\dot{\mu}})$ extending the spinor space. The HS star product in the extended space  is
	\begin{equation}
		(f*g)(Z,Y) = \frac{1}{(2\pi)^4} \int dU dV f(Z+U, Y+U) g(Z-V, Y+V) e^{i U_A V^A}\,. \label{star_product}
	\end{equation}
	Note that for $Z_A$-independent functions it reproduces (\ref{y_star_product}). The following commutation relations  hold true
	\begin{equation}
		[Y_A, Y_B]_* = - [Z_A, Z_B]_* = 2i \epsilon_{AB}, \qquad [Y_A, Z_B]_* = 0\,.
	\end{equation}

	The system of HS equations of \cite{Vasiliev:1992av} is
	\begin{eqnarray}
		&& \d_x W + W*W = 0, \label{e: nonlineareq_1}\\
		&& \d_x S + [W,S]_* = 0\,, \label{e: nonlineareq_2}\\
		&& \d_x B + [W,B]_* = 0\,, \label{e: nonlineareq_3}\\
		&& S*S = i(\theta^A \theta_A + \eta B*\gamma + \bar{\eta} B * \bar{\gamma})\,, \label{e: nonlineareq_4}\\
		&& [S,B]_* = 0\,. \label{e: nonlineareq_5}
	\end{eqnarray}
Here $W(Z, Y; K|x)$ is a one-form that encodes $\omega(Y;K|x)$ while
	$B(Z, Y; K|x)$ is a  zero-form that encodes  $C(Y;K|x)$.
The field  $S(Z, Y; K|x)$ is a space-time zero-form but a one-form in  additional
	differentials $\theta^A = (\theta^\mu, \bar\theta^{\dot{\mu}})$ that anticommute with each other
	and with the space-time de Rham derivative,	
	\begin{equation}
		\{\theta_A, \theta_B\} = \{\theta_A, \d_x\} = 0\,.
	\end{equation}
The central elements of the HS algebra
$\gamma$ and $\bar \gamma$ on the \rhs of (\ref{e: nonlineareq_4}) are
	\begin{equation}\label{e: gamma def}
		\gamma = e^{iz_\mu y^\mu} k \theta^\nu \theta_\nu\,, \qquad \bar{\gamma} = e^{i \bar{z}_{\dot{\mu}} \bar{y}^{\dot{\mu}}} \bar{k} \,  \bar{\theta}^{\dot{\nu}} \,  \bar{\theta}_{\dot{\nu}}\,
	\end{equation}
and $\eta$ is a  free complex phase parameter such that $\eta \,  \bar{\eta} = 1$.
It breaks  parity in the interacting HS theory except for the two cases of $\eta = 1$ or $\eta = i$
	\cite{Sezgin:2003pt}.	
The Klein operators $K = (k, \bar{k})$ satisfy
	\begin{equation}
		\{k,y_\mu\} = \{k,z_\mu\} = 0\,, \qquad [k,\bar{y}_{\dot{\mu}}] = [k,\bar{z}_{\dot{\mu}}] = 0\,, \qquad k^2 = 1\,,
	\end{equation}
	\begin{equation}
	    \left\{\theta_\mu, k\right\}=\left[\bar{\theta}_{\dot{\mu}}, k\right]=0\,, \quad \left[k, \bar{k} \right] = 0 \,,
	\end{equation}
	and analogously for $\bar{k}$. Since  $k^2 = \bar{k}^2=1$, the dependence on $k$ and $\bar k$ is at most bilinear.
The fields decompose into  physical and  topological parts.
 The former are defined by
	\begin{equation}
		W(Z, Y; K|x) = W(Z,Y;-K|x)\,,\qquad
			B(Z, Y; K|x) = -B(Z,Y;-K|x)\,.
	\end{equation}

	\section{Perturbative analysis}\label{PERT}
	
	Equations (\ref{e: verteq_1}), (\ref{e: verteq_2}) can be extracted from nonlinear HS system
(\ref{e: nonlineareq_1})-(\ref{e: nonlineareq_5}) via perturbative expansion. The zero-order
	vacuum solution of the HS equations is
	\begin{align}
		W_0 &= \Omega = \frac{i}{4}\big(\omega_{\mu\nu}(x)y^\mu y^\nu + \bar \omega_{\dot\mu\dot\nu}(x)\bar y^{\dot\mu} \bar y^{\dot\nu} + 2 \lambda h_{\mu \dot\mu}(x) y^{\mu} \bar y^{\dot\mu} \big)\,, \\
		B_0 &= 0\,, \label{B0}\\
		S_0 &= \theta^A Z_A\,.
	\end{align}
	
	The fields $W_0, B_0, S_0$ satisfy equations (\ref{e: nonlineareq_1})-(\ref{e: nonlineareq_5}) if $\omega_{\mu\nu}(x), \bar \omega_{\dot\mu\dot\nu}(x) $ are $AdS_4$ spin-connections and $h_{\mu \dot\mu}(x)$ is $AdS_4$ frame-field. (In the sequel
the inverse $AdS$ radius is set to one,  $\lambda = 1$.) It is important to notice that
	\begin{equation}
		[S_0, f(Z,Y;K)]_* = -2i \theta^A \frac{\partial}{\partial Z^A} f(Z,Y;K) = -2i \d_Z f(Z,Y;K)\,. \label{e: dz}
	\end{equation}
	
	In the first order, equation (\ref{e: nonlineareq_5}) yields
	\begin{equation}
		[S_0, B_1]_* + [S_1, B_0]_* = 0\,.
	\end{equation}
	From (\ref{B0}) and  (\ref{e: dz}) it follows  that $B_1$ is $Z$-independent,
	$B_1 = C(Y;K|x)$. Therefore, eq.(\ref{e: nonlineareq_3}) leads to
	\begin{equation}
		\d_x C(Y;K|x) + [\Omega, C(Y;K|x)]_* = 0\,, \label{dC+CC}
	\end{equation}
that yields (\ref{DC}) in the physical sector. In the sector of topological (auxiliary in terminology of \cite{Vasiliev:1989}) fields, defined as $C(Y;K|x) = C(Y;-K|x)$, equation (\ref{dC+CC}) yields
\begin{equation}
    \bigg(D_L + h^{\mu \dot \mu}\big(y_\mu \bar {\partial}_{\dot {\mu}} + \bar{y}_{\dot{\mu}} \partial_\mu \big)\bigg)C(Y;K|x) = 0\,.
\end{equation}
For topological gauge fields, $\omega(Y;K|x) = - \omega(Y;-K|x)$, the First On-Shell Theorem takes the form \cite{Vasiliev:1992av}
\begin{equation}\label{e: COST TOP}
    R^{top}{ }_{\alpha_1 \ldots \alpha_n, \dot{\beta}_1 \ldots \dot{\beta}_m}(x) =-\bigg[\delta_{0,n} m(m-1) h_{\gamma \dot{\beta}_1} \wedge h^\gamma{}_{\dot{\beta}_2} \bar{C}_{\dot{\beta}_3 \ldots \dot{\beta}_m}(x) + \delta_{0,m} n(n-1) h_{\alpha_1 \dot{\delta}} \wedge h_{\alpha_2}{ }^{\dot{\delta}} C_{\alpha_3 \ldots \alpha_n}(x) \bigg]\,,
\end{equation}
where
\begin{multline}
    R^{top}{ }_{\alpha_1 \ldots \alpha_n, \dot{\beta}_1 \ldots \dot{\beta}_m}(x) = D_L \omega_{\alpha_1 \ldots \alpha_n, \dot{\beta}_1 \ldots \dot{\beta}_m}(x) -  h_{\alpha_1 \dot{\beta}_1}(x)\omega_{\alpha_2 \ldots \alpha_n, \dot{\beta}_2 \ldots \dot{\beta}_m}(x) -\\
    - (n+1) (m+1) h^{\mu \dot{\mu}}(x)\omega_{\mu \alpha_1 \ldots \alpha_n,\dot{\mu} \dot{\beta}_1 \ldots \dot{\beta}_m}(x)\,.
\end{multline}

	The expression for $S_1$ via the field $C$ can be extracted from eq.(\ref{e: nonlineareq_4})
	\begin{equation}\label{e: S1}
		-2i \d_Z S_1 = i\eta C*\gamma + i \bar{\eta} C * \bar{\gamma}\,.
	\end{equation}
	Now we have to solve the differential equation with the exterior differential  $\d_Z$.
	A solution to such an equation  is unique up to the choice of the cohomology class and
its representative.
	Generally,  equation
	\begin{equation}\label{e: generalhomoeq}
		\d_Z f(Z,Y;K;\theta) = g(Z,Y;K;\theta)
	\end{equation}
	with $\d_Z g(Z,Y;K;\theta) = 0$ can be solved by the homotopy trick.
	It can be checked that the \rhs of  (\ref{e: S1}) is $\d_Z$-closed. Firstly,
following \cite{Didenko:2018fgx}, we  choose a nilpotent homotopy operator
	\begin{equation}
		\partial = (Z^A + Q^A) \frac{\partial}{\partial \theta^A}\,,
	\end{equation}
	where
	\begin{equation}
		\frac{\partial Q^B}{\partial Z^A} = 0\,.
	\end{equation}
	Then we introduce operator
	\begin{equation}
		N=\d_{Z} \partial+\partial \d_{Z}=\theta^{A} \frac{\partial}{\partial \theta^{A}}+\left(Z^{A}+Q^{A}\right) \frac{\partial}{\partial Z^{A}}
	\end{equation}
	and the almost inverse operator
	\begin{equation}
		N^{*} g(Z, Y ; \theta):=\int_{0}^{1}  \frac{d t}{t} g(t Z-(1-t) Q, Y ; t \theta), \quad g(-Q , Y ; 0)=0\,.
	\end{equation}
The contracting homotopy operator
	\begin{equation}
		\Delta_{Q}:=\partial N^{*}, \quad \Delta_{Q} g(Z , Y ; \theta)=\left(Z^{A}+Q^{A}\right)
\frac{\partial}{\partial \theta^{A}} \int_{0}^{1} \frac{d t }{t} g(t Z-(1-t) Q , Y ; t \theta)
	\end{equation}
	satisfies the resolution of identity
	\begin{equation}\label{e: resolution of Id}
		\{\d_Z, \Delta_Q \} = 1 - h_Q\,
	\end{equation}
	with $h_Q$ being a cohomology projector
	\begin{equation}
		h_Q f(Z; \theta) = f(-Q; 0)\,.
	\end{equation}
	Hence, resolution of identity yields a particular solution to (\ref{e: generalhomoeq})
	\begin{equation}
		\label{fqg}
		f = \Delta_Q g \,
	\end{equation}
	as long as $h_Q g = 0$, which is true in our case. General solution of (\ref{e: generalhomoeq}) is
	\begin{equation}
		f(Z,Y;\theta) = \Delta_Q g(Z,Y;\theta) + h(Y) + \d_Z \epsilon(Z,Y;\theta)\,,
	\end{equation}
	where $h(Y)$ is a cohomology representative and $\epsilon(Z,Y;\theta)$ is a parameter
	of gauge transformation ($\d_Z$-exact term). Transition from one $Q$ to another affects
	the $h$ and $\epsilon$-dependent parts of the solution.
	The choice of  $Q$ in (\ref{fqg}) affects
	the choice of field variables, that can be essential for the analysis of locality.
	Originally the choice of $Q = 0$ known as the conventional homotopy was studied \cite{Vasiliev:1992av},
	which led to the First On-Shell Theorem. More complex shifts were applied in
	\cite{Gelfond:2018vmi}-\cite{Gelfond:2021two} for the analysis of locality problem in
the non-linear HS theory.
	
	\section{Shifted homotopy}\label{Shifted homotopies}
	
	For the subsequent analysis we recall, following \cite{Didenko:2018fgx},
	some properties of the operators $\Delta_Q$ and $h_Q$ defined in the previous section.
	Firstly, operators $\Delta_Q$ and $\Delta_P$
	anticommute
	\begin{equation}
		\Delta_Q \Delta_P = - \Delta_P \Delta_Q\,.
	\end{equation}
	Analogously,
	\begin{equation}
		h_P \Delta_Q = - h_Q \Delta_P\,.
	\end{equation}
	
Confining ourselves to the holomorphic variables $(Z_A, Y_A, K) \rightarrow (z_\mu, y_\mu, k)$,
let us write down how $\Delta_b \Delta_a$ and $h_c \Delta_b \Delta_a$ act
	\begin{equation}
		\Delta_b \Delta_a f(z,y) \theta^\mu \theta_\mu = 2 \int_{[0,1]^3} d^3 \tau \delta(1-\tau_1 -
\tau_2 - \tau_3) (z+b)_\nu (z+a)^\nu f(\tau_1 z - \tau_3 b -\tau_2 a, y)\,, \label{dd}
	\end{equation}
	\begin{equation}
		h_c \Delta_b \Delta_a f(z,y) \theta^\mu \theta_\mu = 2 \int_{[0,1]^3} d^3 \tau \delta(1-
\tau_1 - \tau_2 - \tau_3) (b-c)_\nu (a-c)^\nu f(-\tau_1 c - \tau_3 b - \tau_2 a, y)\, \label{hdd}.
	\end{equation}
	
	Note that from (\ref{hdd}) it follows that for any parameter $\kappa$
	\begin{equation}
		h_{(\kappa + 1)q_2 - \kappa q_1} \Delta_{q_2} \Delta_{q_1} = 0 \,. \label{hdd=0}
	\end{equation}
    This identity will have important implications later on.

	Application of formulas (\ref{dd}), (\ref{hdd})  to $\gamma$ yields
	\begin{eqnarray}
		&& \Delta_b \Delta_a \gamma = 2 \int_{[0,1]^3} d^3 \tau \delta(1-\tau_1 - \tau_2 - \tau_3) (z+b)_\nu (z+a)^\nu e^{i(\tau_1 z - \tau_2 a - \tau_3 b)_\mu y^\mu} k\,,\\
		&& h_c \Delta_b \Delta_a \gamma = 2 \int_{[0,1]^3} d^3 \tau \delta(1-\tau_1 - \tau_2 - \tau_3)
 (b-c)_\nu (a-c)^\nu e^{-i(\tau_1 c + \tau_2 a + \tau_3 b)_\mu y^\mu} k \,.\label{e: hdd_int}
	\end{eqnarray}
	
	Yet another important property of the operators $\Delta_Q$ and $h_P$,
implying the $z$-independence of the vertices resulting from equations
(\ref{e: nonlineareq_1})--(\ref{e: nonlineareq_5}), is
	\begin{equation}\label{e: hdd_0}
		\left(\Delta_{d}-\Delta_{c}\right)\left(\Delta_{a}-\Delta_{b}\right) \gamma=\left(h_{d}-h_{c}\right) \Delta_{a} \Delta_{b} \gamma\,.
	\end{equation}
		It has a consequence
	\begin{equation}\label{e: hdd}
		(\Delta_c \Delta_b - \Delta_c \Delta_a + \Delta_b \Delta_a) \gamma = h_c \Delta_b \Delta_a \gamma\,.
	\end{equation}
	
	Other remarkable properties of the shifted homotopy operators  also obtained in \cite{Didenko:2018fgx}
are the so-called star-exchange relations
	with $z$-independent elements
	\begin{equation}
		\Delta_{q+\alpha y} ( C(y;k) * \phi(z,y;k;\theta)) = C(y;k) * \Delta_{q+(1-\alpha)p + \alpha y} \phi(z,y;k;\theta), \label{e: Cexchange}
	\end{equation}
	\begin{equation}
		\Delta_{q+\alpha y} ( \phi(z,y;k;\theta) * k^m * C(y;k) ) = \Delta_{q+(-1)^m (1-\alpha)p + \alpha y} (\phi(z,y;k;\theta)) *k^m * C(y;k) \label{e: exchangeC}\,.
	\end{equation}
	Here
	\begin{equation}
		p_\mu C(Y;K) = C(Y;K)p_\mu := -i \frac{\partial}{\partial y^\mu} C(Y;K)\,.
	\end{equation}
	Also, for the central element $\gamma$,
	\begin{equation}\label{e: gammaexch}
		\Delta_{q} \gamma * C(y;k) = C(y;k) * \Delta_{q+2p} \gamma\,.
	\end{equation}
	Analogous properties  hold true for the cohomology projector $h_q$	
	\begin{equation}
		h_{q+\alpha y} (C(y;k) * \phi(z,y;k;\theta)) = C(y;k) * h_{q+(1-\alpha)p+\alpha y} \phi(z,y;k;\theta),
	\end{equation}
	\begin{equation}
		h_{q+\alpha y} ( \phi(z,y;k;\theta) * k^m * C(y;k) ) = h_{q+(-1)^m (1-\alpha)p + \alpha y} (\phi(z,y;k;\theta)) *k^m * C(y;k)\,.
	\end{equation}

	\section{General shift parameters }\label{Mod Shift}
	
	In this section we calculate $\Upsilon(\omega, \omega, C)$ vertex using the shifted
	homotopy operators involving shift parameters that act on
	the argument of $\omega$. We adopt notation from \cite{Didenko:2018fgx}
	\begin{equation}
		t_\mu \omega (Y;K|x) = -i \frac{\partial}{\partial y^\mu} \omega(Y;K|x)\,.
	\end{equation}
	
	Let us start with equation (\ref{e: S1}). We choose a solution (\ref{fqg}) with
$Q^\mu = q^\mu + \alpha y^\mu + \lambda p^\mu$ where $q^\mu$ and $\alpha, \lambda$  are free constants. Then, for $S_1 = S_1^\eta +S_1^{\bar\eta}$, in the $\bar\eta$-independent (holomorphic)
sector we obtain
	\begin{equation}\label{e: S1_solution}
		S_1^\eta = -\frac{\eta}{2} \Delta_{q + \alpha y + \lambda p} \big(C * \gamma \big)\,.
	\end{equation}
	The next step is to solve eq.(\ref{e: nonlineareq_2})  which yields in the first order
	\begin{equation}\label{e: W1}
\d_z W_1^\eta = -\frac{i}{2}\big(\d_x S^\eta_1 + \omega * S_1^\eta + S_1^\eta * \omega \big)\,.
	\end{equation}
	
	Equation (\ref{e: W1}) decomposes  into two subsystems. This is because,
as pointed out in \cite{Vasiliev:1989}, HS unfolded equations remain consistent with the
fields $\omega$ and $C$ valued in any associative algebra which implies that
they are associated with the so-called
$A_\infty$-algebra \cite{10.2307/1993608, 10.2307/1993609}.
From this it follows, that the following equations have to be separately satisfied
	\begin{align}
		\d_z W_1^{\eta (1)} &= -\frac{i}{2}\big(\d_x S^\eta_1\big|_{\omega*C} + \omega * S^\eta_1 \big)\,, \label{e: W11}\\
		\d_z W_1^{\eta (2)} &= -\frac{i}{2}\big(\d_x S^\eta_1\big|_{C*\omega} + S_1^\eta * \omega \big) \label{e: W12}\,.
	\end{align}
	Indeed,  using equation (\ref{e: nonlineareq_3})
equations (\ref{e: W11}) and (\ref{e: W12}) can be checked to separately
satisfy consistency conditions. While doing so, it is important to remember
 that in the term $\d_x \d_z S_1^\eta$
 one must keep only the term with the chosen  order of $\omega$ and $C$
 resulting from equation (\ref{dC+CC}).
	
	Hence, we can apply independent shifts for the different components of $W_1^\eta$.
Let us choose the following solutions to Eqs.~(\ref{e: W11}) and (\ref{e: W12})
with $t$-dependent shifts $Q_i^\mu= l_i^\mu + n_i t^\mu + \beta_i y^\mu$. As shown in \cite{Didenko:2018fgx}, uniform shifts $\Delta_{\gamma(p + y)}$ in both $S$ and $W$ do not affect the form of the vertices. The freedom in the uniform shifts allows us to fix the $p$ shift for $W_1$ to zero, so this is in fact the most general form of a linear shift for this set of variables.
In general, both orderings in $W_1$ must result from  the same homotopy procedure. However, one can start with introducing different shifts $n_i t$ and $\beta_i y$. Any $n_i$ respect the compatibility conditions independently, while, as we show later on, $\beta_i y$ shifts have to vanish.
 So,
	\begin{align}
		W^{\eta (1)}_1 &= \frac{1}{2i} \Delta_{l_1 + n_1 t + \beta_1 y }\big(\d_x S^\eta_1\big|_{\omega*C} + \omega * S^\eta_1 \big) \,,\label{e: W11tilde} \\
		W^{\eta (2)}_1 &= \frac{1}{2i} \Delta_{l_2 + n_2 t +\beta_2  y }\big(\d_x S^\eta_1\big|_{C*\omega} + S_1^\eta * \omega \big) \,\label{e: W12tilde}
	\end{align}
	with $l_{i}^\mu$ and $n_i$ being some constants.
 Plugging in (\ref{e: S1_solution}) and applying star-exchange
 formulae (\ref{e: Cexchange}), (\ref{e: exchangeC}), (\ref{e: gammaexch}) we obtain
	\begin{align}
		W^{\eta (1)}_1 &= \frac{\eta}{4i} \omega*C*\Delta_{l_1 + n_1 t + \beta_1 y + (1 - \beta_1)(t+p)}\big(\Delta_{\tilde{q} + (1-\alpha + \lambda)p} - \Delta_{\tilde{q} + (1-\alpha + \lambda)(t+p)}\big)\gamma\,, \label{e: W11new}\\
		W^{\eta (2)}_1 &= \frac{\eta}{4i} C*\omega*\Delta_{l_2 + n_2 t + \beta_2 y + (1-\beta_2)(p+t)}\big(\Delta_{\tilde{q} + (1-\alpha + \lambda)(t+p)} - \Delta_{\tilde{q} + (1-\alpha + \lambda)p + 2t}\big)\gamma \label{e: W12new}\,,
	\end{align}
	where $\tilde{q} = q + \alpha y$.

	Now consider equation (\ref{e: nonlineareq_1}). In the first order it yields
	\begin{equation}
		\d \omega + \omega*\omega + \d W_1^\eta + \omega*W_1^\eta + W_1^\eta *\omega = 0\,,
	\end{equation}
	where $W_1^\eta = W_1^{\eta(1)} + W_1^{\eta(2)}$. Using (\ref{e: W11new}) and (\ref{e: W12new}) and applying formulae (\ref{e: hdd}),(\ref{e: Cexchange}),(\ref{e: exchangeC}),(\ref{e: gammaexch}) one can obtain
	\begin{equation}
		\d \omega + \omega*\omega + \Upsilon^\eta(\omega,\omega,C)  + \Upsilon^\eta(C,\omega,\omega) + \Upsilon^\eta(\omega,C,\omega) = 0\,.
	\end{equation}
	Direct calculation of the vertices yields
	\begin{multline}\label{e: full_vert_beginning}
		\Upsilon^\eta(\omega,\omega,C) = \frac{\eta}{4i} \omega*\omega*C*\big[h_{l_1 + n_1(t_1 + t_2) + \beta_1 y + (1 - \beta_1)(p + t_1 + t_2)}\Delta_{\tilde{q} + (1-\alpha + \lambda)(t_1+t_2+p)}\Delta_{\tilde{q} + (1-\alpha + \lambda)p}\gamma + \\
		+ h_{l_1 + n_1t_1 + \beta_1 y + (1-\beta_1)(p+t_1+t_2)}\Delta_{\tilde{q} + (1-\alpha + \lambda)(t_2+p)}\Delta_{\tilde{q} + (1-\alpha + \lambda)(t_1 + t_2 +p)}\gamma + \\
		+ h_{l_1 + n_1 t_2 + \beta_1 y + (1-\beta_1)(p+t_2)}\Delta_{\tilde{q} + (1-\alpha + \lambda)p}\Delta_{\tilde{q} + (1-\alpha + \lambda)(t_2 + p)}\gamma + \\
  + h_{\tilde{q} + (1-\alpha + \lambda)p}\Delta_{\tilde{q} + (1-\alpha + \lambda)(t_1+t_2+p)}\Delta_{\tilde{q} + (1-\alpha + \lambda)(t_2 + p)}\gamma\big]\,,
	\end{multline}
		\begin{multline}\label{e: full_vert_ending}
		\Upsilon^\eta(C,\omega,\omega) = \frac{\eta}{4i} C*\omega*\omega*\big[h_{l_2 + n_2t_2 + \beta_2 y + (1-\beta_2)(p+t_1+t_2)}\Delta_{\tilde{q} + (1-\alpha + \lambda)(t_1+t_2+p)}\Delta_{\tilde{q}+(1-\alpha + \lambda)(t_1+p) + 2t_2}\gamma + \\
		+ h_{l_2 + n_2(t_1+t_2) + \beta_2 y + (1-\beta_2)(p+t_1+t_2)}\Delta_{\tilde{q}+(1-\alpha + \lambda)p +2t_1 +2t_2}\Delta_{\tilde{q} + (1-\alpha + \lambda)(t_1+t_2+p)}\gamma + \\
		+ h_{l_2 + n_2t_1 + \beta_2 y + (1-\beta_2)(p+t_1) + 2t_2}\Delta_{\tilde{q} + (1-\alpha + \lambda)(t_1+p) + 2t_2}\Delta_{\tilde{q} + (1-\alpha + \lambda)p +2t_1 +2t_2}\gamma + \\
		+ h_{\tilde{q} + (1-\alpha + \lambda)p +2t_1 +2t_2}\Delta_{\tilde{q} + (1-\alpha + \lambda)(t_1+p) + 2t_2}\Delta_{\tilde{q} + (1-\alpha + \lambda)(t_1+t_2+p)}\gamma\big]\,,
	\end{multline}

	\begin{multline}
		\Upsilon^\eta(\omega,C,\omega) = \frac{\eta}{4i} \omega*C*\omega*\big[h_{\tilde{q} + (1-\alpha )(t_1+t_2+p)}\Delta_{\tilde{q} + (1-\alpha +\lambda)(t_1+p) + 2t_2}\Delta_{\tilde{q} + (1-\alpha +\lambda )(t_2+p)}\gamma + \\
		+ h_{\tilde{q} + (1-\alpha + \lambda )(t_1+p) + 2t_2}\Delta_{\tilde{q} + (1-\alpha +\lambda)p + 2t_2}\Delta_{\tilde{q}+(1-\alpha +\lambda )(t_2+p)}\gamma + \\
        + h_{l_1 + n_1 t_1 + \beta_1 y + (1-\beta_1)(t_1+t_2+p)}\Delta_{\tilde{q} + (1-\alpha +\lambda)(t_1+t_2+p)}\Delta_{\tilde{q} + (1-\alpha +\lambda)(t_2+p)}\gamma + \\
        + h_{l_2 + n_2 t_2 + \beta_2 y + (1-\beta_2)(p+t_1+t_2)}\Delta_{\tilde{q} + (1-\alpha +\lambda)(t_1+p)+2t_2}\Delta_{\tilde{q} + (1-\alpha + \lambda)(t_1+t_2+p)}\gamma + \\
		+ h_{l_1 + n_1 t_1 + \beta_1 y + (1-\beta_1)(p+t_1) + 2t_2}\Delta_{\tilde{q} + (1-\alpha+\lambda)p + 2t_2}\Delta_{\tilde{q}+(1-\alpha+\lambda)(t_1+p) + 2t_2}\gamma + \\
		+ h_{l_2 + n_2t_2 + \beta_2 y + (1-\beta_2)(p+t_2)}\Delta_{\tilde{q} + (1-\alpha+\lambda)(t_2+p)}\Delta_{\tilde{q} + (1-\alpha +\lambda)p + 2t_2}\gamma \big]\,.
	\end{multline}
	
	To simplify  further analysis we set $l_i^\mu = q^\mu = 0$, which
	is anyway necessary   since  non-zero constant spinors like
$l_{i}^\mu$ and $q^\mu$  violate Lorentz covariance. In practice, the presence of such constant parameters
would result in terms containing $\omega h$ in the vertices so that
the Lorentz connection would not enter solely via the Lorentz covariant
derivative.

Now we use (\ref{e: hdd_int}) and	evaluate star products in the vertices (\ref{e: full_vert_beginning})-(\ref{e: full_vert_ending}) using  the following
notations for the argument of the exponent
	\begin{align}
		\varkappa_{\omega \omega C}(y,t_i,p) &= y^\mu (t_1+t_2+p)_\mu + t_1^\mu t_2{}_\mu + (t_1+t_2)^\mu p_\mu\,, \\
		\varkappa_{\omega C \omega}(y,t_i,p) &= y^\mu (t_1+t_2+p)_\mu + t_1^\mu p_\mu + (t_1+p)^\mu t_2{}_\mu\,, \\
		\varkappa_{C \omega \omega}(y,t_i,p) &= y^\mu (t_1+t_2+p)_\mu + p^\mu t_1{}_\mu + (t_1+p)^\mu t_2{}_\mu\,.
	\end{align}
	This yields

	1) $\omega*\omega*C$-terms
	\begin{multline}\label{e: preexp_first}
		\omega*\omega*C*h_{n_1(t_1 + t_2) + \beta_1 y + (1-\beta_1)(p+t_1+t_2)}\Delta_{\alpha y + (1-\alpha+\lambda)(t_1+t_2+p)}\Delta_{\alpha y + (1-\alpha+\lambda)p}\gamma = \\ = 2 \omega \bar{*} \omega \bar{*} C \int_{[0,1]^3} d^3\tau_i \delta(1-\sum_{i=1}^{3}\tau_i)  (1 - \alpha+\lambda) \Bigg[(\alpha-\beta_1)y^\mu (t_1+t_2)_\mu - \lambda p_\mu (t_1 + t_2)^\mu \Bigg]\exp{i \varkappa_{\omega \omega C}(y,t_i,p)} \\ \text{exp}\bigg[-i (y + t_1 + t_2 + p)^\nu \big(\tau_1 [n_1(t_1 + t_2) + (1-\beta_1)(p+t_1+t_2)]  + \tau_2 [(1-\alpha+\lambda)p] + \tau_3 [(1-\alpha+\lambda)(t_1+t_2+p)]\big)_\nu \bigg]k\,,
	\end{multline}
	where $\bar *$ is the star product in the antiholomorphic variables $\bar y_{\dot \mu}$
	\begin{equation}
		f(\bar{y})\bar{*}g(\bar{y}) = f(\bar{y}) e^{i {\epsilon}^{\dot{\mu} \dot{\nu}}\overleftarrow{\bar{\partial}}_{\dot{\mu}} \overrightarrow{\bar{\partial}}_{\dot{\nu}} } g(\bar{y})\,, \label{bar_star_product}
	\end{equation}
	\begin{multline}
		\omega*\omega*C*h_{n_1 t_1 + \beta_1 y + (1-\beta_1)(p+t_1+t_2)}\Delta_{\alpha y + (1-\alpha+\lambda)(t_2+p)}\Delta_{\alpha y + (1-\alpha+\lambda)(t_1 + t_2 +p)}\gamma = \\ = 2 \omega \bar{*} \omega \bar{*} C \int_{[0,1]^3} d^3\tau_i \delta(1-\sum_{i=1}^{3}\tau_i) (\alpha - 1-\lambda)\Bigg[(\alpha-\beta_1)y^\mu t_1{}_\mu - \lambda(p+t_2)_\mu t_1^\mu \Bigg]  \exp{i \varkappa_{\omega \omega C}(y,t_i,p)} \\ \text{exp}\bigg[-i (y + t_1 + t_2 + p)^\nu \big(\tau_1 [n_1 t_1 + (1-\beta_1)(p+t_1+t_2)] +\tau_2 [(1-\alpha+\lambda)(t_1 + t_2 +p)] + \tau_3 [(1-\alpha+\lambda)(t_2+p)]\big)_\nu \bigg] k\,,
	\end{multline}
	\begin{multline}
		\omega*\omega*C*h_{n_1 t_2 + \beta_1 y + (1-\beta_1)(p+t_2)}\Delta_{\alpha y + (1-\alpha+\lambda)p}\Delta_{\alpha y + (1-\alpha+\lambda)(t_2 + p)}\gamma = \\ = 2 \omega \bar{*} \omega \bar{*} C \int_{[0,1]^3} d^3\tau_i \delta(1-\sum_{i=1}^{3}\tau_i)(\alpha - 1-\lambda)\Bigg[(\alpha-\beta_1)(y+t_1)^\mu t_2{}_\mu - \lambda p_\mu t_2^\mu \Bigg] \exp{i \varkappa_{\omega \omega C}(y,t_i,p)} \\ \text{exp}\bigg[-i (y + t_1 + t_2 + p)^\nu \big(\tau_1 [n_1 t_2 + (1-\beta_1)(p+t_2)] +\tau_2 [(1-\alpha+\lambda)(t_2 + p)] + \tau_3 [(1-\alpha + \lambda)p]\big)_\nu \bigg] k\,,
	\end{multline}
	\begin{multline}
		\omega*\omega*C*h_{\alpha y + (1-\alpha + \lambda)p}\Delta_{\alpha y + (1-\alpha + \lambda)(t_1+t_2+p)}\Delta_{\alpha y + (1-\alpha + \lambda)(t_2 + p)}\gamma = \\ = 2 \omega \bar{*} \omega \bar{*} C \int_{[0,1]^3} d^3\tau_i \delta(1-\sum_{i=1}^{3}\tau_i)( \alpha - 1 - \lambda)^2 t_2{}^\mu t_1{}_\mu \exp{i \varkappa_{\omega \omega C}(y,t_i,p)} \\ \text{exp}\bigg[-i (y + t_1 + t_2 + p)^\nu \big(\tau_1 [(1-\alpha + \lambda)p] +\tau_2 [(1-\alpha + \lambda)(t_2 + p)] + \tau_3 [(1-\alpha + \lambda)(t_1+t_2+p)] \big)_\nu \bigg] k\,.
	\end{multline}
	
	2) $\omega*C*\omega$-terms
	\begin{multline}
		\omega*C*\omega*h_{\alpha y + (1-\alpha+\lambda)(t_1+t_2+p)}\Delta_{\alpha y + (1-\alpha+\lambda)(t_1+p) + 2t_2}\Delta_{\alpha y + (1-\alpha+\lambda)(t_2+p)}\gamma = \\
		= 2 \omega \bar{*} C\bar{*}\omega \int_{[0,1]^3} d^3\tau_i \delta(1-\sum_{i=1}^{3}\tau_i)(1+\alpha-\lambda)(\alpha-1-\lambda) t_1{}^\mu t_2{}_\mu \exp{i \varkappa_{\omega C \omega}(y,t_i,p)} \\  \text{exp} \bigg[-i (y + t_1 + t_2 + p)^\nu \big(\tau_1 [(1-\alpha+\lambda)(t_1+t_2+p)] + \tau_2 [(1-\alpha+\lambda)(t_2+p)] + \tau_3 [(1-\alpha+\lambda)(t_1+p) + 2t_2] \big)_\nu \bigg] k  \,,
	\end{multline}
	\begin{multline}
		\omega*C*\omega*h_{\alpha y + (1-\alpha+\lambda)(t_1+p) + 2t_2}\Delta_{\alpha y + (1-\alpha+\lambda)p + 2t_2}\Delta_{\alpha y+(1-\alpha+\lambda)(t_2+p)}\gamma = \\ = 2\omega \bar{*} C\bar{*}\omega\int_{[0,1]^3} d^3\tau_i \delta(1-\sum_{i=1}^{3}\tau_i)(\alpha+1-\lambda)(\alpha-1-\lambda) t_1{}^\mu t_2{}_\mu \exp{i \varkappa_{\omega C \omega}(y,t_i,p)} \\ \text{exp}\bigg[-i (y + t_1 + t_2 + p)^\nu \big(\tau_1 [(1-\alpha+\lambda)(t_1+p) +  2t_2] +\tau_2 [(1-\alpha+\lambda)(t_2+p)] + \tau_3 [(1-\alpha+\lambda)p + 2t_2] \big)_\nu \bigg] k  \,,
	\end{multline}
	\begin{multline}
		\omega*C*\omega*h_{n_1 t_1 + \beta_1 y + (1-\beta_1)(p+t_1+t_2)}\Delta_{\alpha y + (1-\alpha+\lambda)(t_1+t_2+p)}\Delta_{\alpha y + (1-\alpha+\lambda)(t_2+p)}\gamma = \\ = - 2\omega \bar{*} C\bar{*}\omega\int_{[0,1]^3} d^3\tau_i \delta(1-\sum_{i=1}^{3}\tau_i)(\alpha-1-\lambda)\Bigg[(\alpha - \beta_1) y{}^\mu t_1{}_\mu + \lambda (p + t_2)^\mu t_1{}_\mu\Bigg] \exp{i \varkappa_{\omega C \omega}(y,t_i,p)} \\ \text{exp}\bigg[-i (y + t_1 + t_2 + p)^\nu \big(\tau_1 [n_1 t_1 + (1-\beta_1)(p+t_1+t_2)]  +\tau_2 [(1-\alpha+\lambda)(t_2+p)] + \tau_3 [(1-\alpha+\lambda)(t_1+t_2+p)] \big)_\nu \bigg] k  \,,
	\end{multline}
	\begin{multline}
		\omega*C*\omega*h_{n_2 t_2 + \beta_2 y + (1-\beta_2)(p+t_1+t_2)}\Delta_{\alpha y + (1-\alpha+\lambda)(t_1+p)+2t_2}\Delta_{\alpha y + (1-\alpha+\lambda)(t_1+t_2+p)}\gamma = \\ = 2\omega \bar{*} C\bar{*}\omega\int_{[0,1]^3} d^3\tau_i \delta(1-\sum_{i=1}^{3}\tau_i)(\alpha + 1 - \lambda)\bigg[(\alpha - \beta_2) y{}^\mu t_2{}_\mu + \lambda (p + t_1)^\mu t_2{}_\mu\bigg] \exp{i \varkappa_{\omega C \omega}(y,t_i,p)} \\ \text{exp}\bigg[-i (y + t_1 + t_2 + p)^\nu \big(\tau_1 [n_2 t_2 + (1-\beta_2)(t_1 + t_2 + p)] +\tau_2 [(1-\alpha+\lambda)(t_1+t_2+p)] + \\ +\tau_3 [(1-\alpha+\lambda)(t_1+p)+2t_2] \big)_\nu \bigg] k  \,,
	\end{multline}
	\begin{multline}
		\omega*C*\omega*h_{n_1 t_1 + \beta_1 y + (1-\beta_1)(p+t_1) + 2t_2}\Delta_{\alpha y + (1-\alpha+\lambda)p + 2t_2}\Delta_{\alpha y+(1-\alpha+\lambda)(t_1+p) + 2t_2}\gamma = \\ = 2\omega \bar{*} C\bar{*}\omega \int_{[0,1]^3} d^3\tau_i \delta(1-\sum_{i=1}^{3}\tau_i)(\alpha - 1 - \lambda)\bigg[(\alpha - \beta_1) (y{} + t_2{})^\mu t_1{}_\mu + \lambda p^\mu t_1{}_\mu\bigg]\exp{i \varkappa_{\omega C \omega}(y,t_i,p)} \\ \text{exp}\bigg[-i (y + t_1 + t_2 + p)^\nu \big(\tau_1 [n_1 t_1 + (1-\beta_1)(p+t_1) + 2t_2]  +\tau_2 [(1-\alpha+\lambda)p + 2t_2] + \\ + \tau_3 [(1-\alpha+\lambda)(t_1+p) + 2t_2] \big)_\nu \bigg] k \,,
	\end{multline}
    \begin{multline}
		\omega*C*\omega*h_{n_2 t_2 + \beta_2 y + (1-\beta_2)(p+t_2)}\Delta_{\alpha y + (1-\alpha+\lambda)(t_2+p)}\Delta_{\alpha y + (1-\alpha+\lambda)p + 2t_2}\gamma = \\ = -2\omega \bar{*} C\bar{*}\omega \int_{[0,1]^3} d^3\tau_i \delta(1-\sum_{i=1}^{3}\tau_i)(\alpha + 1 - \lambda)\bigg[(\alpha - \beta_2)(y{} + t_1{})^\mu t_2{}_\mu + \lambda p^\mu t_2{}_\mu\bigg]\exp{i \varkappa_{\omega C \omega}(y,t_i,p)} \\ \text{exp}\bigg[-i (y + t_1 + t_2 + p)^\nu \big(\tau_1 [n_2 t_2 + (1-\beta_2)(p+t_2)] +\tau_2 [(1-\alpha+\lambda)p + 2t_2] + \tau_3 [(1-\alpha+\lambda)(t_2+p)] \big)_\nu \bigg] k  \,.
    \end{multline}
	
	3) $C*\omega*\omega$-terms
	
	\begin{multline}
		C*\omega*\omega*h_{n_2t_2 + \beta_2 y + (1-\beta_2)(t_1 + t_2 + p)}\Delta_{\alpha y + (1-\alpha+\lambda)(t_1+t_2+p)}\Delta_{\alpha y+(1-\alpha + \lambda)(t_1+p) + 2t_2}\gamma = \\ = -2 C \bar{*} \omega \bar{*} \omega\int_{[0,1]^3} d^3\tau_i \delta(1-\sum_{i=1}^{3}\tau_i)(\alpha + 1-\lambda)\Bigg[(\alpha-\beta_2)y^\mu t_2{}_\mu - \lambda(p+t_1)_\mu t_2{}^\mu \Bigg] \exp{i \varkappa_{C \omega \omega}(y,t_i,p)} \\ \text{exp}\bigg[-i (y + t_1 + t_2 + p)^\nu \big(\tau_1 [n_2t_2 + (1-\beta_2)(t_1 + t_2 + p)]  +\tau_2 [(1-\alpha+\lambda)(t_1+p) + 2t_2] + \\ + \tau_3 [(1-\alpha+\lambda)(t_1+t_2+p)] \big)_\nu \bigg] k \,,
	\end{multline}
	\begin{multline}
		C*\omega*\omega*h_{n_2(t_1+t_2) + \beta_2 y + (1-\beta_2)(p + t_1 + t_2)}\Delta_{\alpha y+(1-\alpha+\lambda)p +2t_1 +2t_2}\Delta_{\alpha y + (1-\alpha+\lambda)(t_1+t_2+p)}\gamma = \\ = 2 C \bar{*} \omega \bar{*} \omega\int_{[0,1]^3} d^3\tau_i \delta(1-\sum_{i=1}^{3}\tau_i)(\alpha + 1-\lambda)\Bigg[(\alpha-\beta_2)y^\mu (t_1+t_2)_\mu - \lambda p_\mu (t_1 + t_2)^\mu \Bigg] \exp{i \varkappa_{C \omega \omega}(y,t_i,p)} \\ \text{exp}\bigg[-i (y + t_1 + t_2 + p)^\nu \big(\tau_1 [n_2(t_1+t_2) + (1-\beta_2)(p+t_1+t_2)] +\tau_2 [(1-\alpha+\lambda)(t_1+t_2+p)] + \\ + \tau_3 [(1-\alpha+\lambda)p +2t_1 +2t_2] \big)_\nu \bigg] k \,,
	\end{multline}
	\begin{multline}
		C*\omega*\omega*h_{n_2 t_1 + \beta_2 y + (1-\beta_2)(p+t_1) + 2t_2}\Delta_{\alpha y + (1-\alpha+\lambda)(t_1+p) + 2t_2}\Delta_{\alpha y + (1-\alpha+\lambda)p +2t_1 +2t_2}\gamma = \\ = -2 C \bar{*} \omega \bar{*} \omega\int_{[0,1]^3} d^3\tau_i \delta(1-\sum_{i=1}^{3}\tau_i)(\alpha + 1-\lambda)\Bigg[(\alpha-\beta_2)(y+t_2)^\mu t_1{}_\mu - \lambda p_\mu t_1{}^\mu \Bigg] \exp{i \varkappa_{C \omega \omega}(y,t_i,p)} \\ \text{exp}\bigg[-i (y + t_1 + t_2 + p)^\nu \big(\tau_1 [n_2 t_1 + (1-\beta_2)(p+t_1) + 2t_2 ] + \tau_2 [(1-\alpha+\lambda)p +2t_1 +2t_2] + \\ +\tau_3 [(1-\alpha + \lambda)(t_1+p) + 2t_2] \big)_\nu \bigg] k \,,
	\end{multline}
	\begin{multline}\label{e: preexp_last}
		C*\omega*\omega*h_{\alpha y + (1-\alpha+\lambda)p +2t_1 +2t_2}\Delta_{\alpha y + (1-\alpha+\lambda)(t_1+p) + 2t_2}\Delta_{\alpha y + (1-\alpha+\lambda)(t_1+t_2+p)}\gamma = \\ = 2 C \bar{*} \omega \bar{*} \omega\int_{[0,1]^3} d^3\tau_i \delta(1-\sum_{i=1}^{3}\tau_i)(1 + \alpha - \lambda)^2 t_2^\mu t_1{}_\mu \exp{i \varkappa_{C \omega \omega}(y,t_i,p)} \\ \text{exp}\bigg[-i (y + t_1 + t_2 + p)^\nu \big(\tau_1 [(1-\alpha+\lambda)p +2t_1 +2t_2] +\tau_2 [(1-\alpha+\lambda)(t_1+t_2+p)] + \tau_3 [(1-\alpha+\lambda)(t_1+p) + 2t_2] \big)_\nu \bigg] k \,.
	\end{multline}

    One can notice similarities in different vertices resulting from the antiautomorphism
    $\rho$  of the
HS star-product algebra,
\begin{equation}
		\rho\bigg( f(Z,Y;K;\theta) \bigg) = f(-i Z, i Y; K; -i \theta)\,,
	\end{equation}
that leaves invariant  non-linear HS equations (\ref{e: nonlineareq_1})-(\ref{e: nonlineareq_5}) \cite{Vasiliev:1999ba}.
Indeed
it is easy to see that  application of such antiautomorphism along with the substitution  $\alpha \leftrightarrow -\alpha$, $t_1 \leftrightarrow t_2$ $n_1 \leftrightarrow -n_2$, $\lambda \leftrightarrow -\lambda$ and $\beta_1 \leftrightarrow -\beta_2$ maps some  pairs of terms to each other.
Namely, the terms of the vertex $\Upsilon^\eta(\omega,\omega,C)$
are mapped  to those of $\Upsilon^\eta(C,\omega,\omega)$, while a half of the terms in
$\Upsilon^\eta(\omega, C, \omega)$ is mapped to the other half.

	\section{Admissible shift parameters}\label{Adm Param}

	As explained above,
	 non-zero constant spinors $q^\mu$ or $l_{i}^\mu$ manifestly violate Lorentz invariance and hence are not allowed. The analysis of the role of the parameters $\alpha$, $\lambda$ and $\beta_i$ requires a bit more work. To respect the form of First On-Shell Theorem
	for the AdS background $\omega = \Omega$  vertices  should have the $y$-independent form
	$h_\mu^{\ \dot\mu} h^{\mu\dot\nu}\,\bar\dd_{\dot\mu}\bar\dd_{\dot\nu} C(0,\bar y\, |x)$ or $h_\mu^{\ \dot\mu} h^{\mu\dot\nu}\,\bar y_{\dot\mu}\bar y_{\dot\nu} C(0,\bar y\, |x)$
in the $\eta$-sector.
	Therefore, it is instrumental to analyse the  $y$-dependence
	of the $C$-field in the vertex. To this end let us inspect all results of multiplication of two fields $\Omega$ and a single field $C$
paying attention to the terms of the form $hh\bar{\partial}\bar{\partial}C$.
Recall that in the previous analysis, arguments of both $\Omega$ and $C$ were
 uplifted into a single exponent by virtue of the Taylor formula
	\begin{equation}
		f(a) = \exp(a \frac{\d}{\d b})f(b)\bigg|_{b = 0}\,
	\end{equation}
	with an auxiliary variable $b$. Proceeding this way,
let us assign the auxiliary variables $y_1$ and $y_2$ to the first and second factors of
$\Omega$ in the ordered product, respectively. We will use the fact that a product of two frame fields can be decomposed into  irreducible parts as
\begin{equation}\label{e: HH_decomp}
    h^{\nu\dot\nu} h^{\lambda\dot\lambda} = \frac{1}{2}H^{\nu\lambda}\epsilon^{\dot\nu\dot\lambda}+\frac{1}{2}\bar H^{\dot\nu\dot\lambda}\epsilon^{\nu\lambda},
\end{equation}
where
\begin{align}
    H^{\nu\lambda}=H^{(\nu\lambda)}:=h^{\nu}_{\ \dot\gamma} h^{\lambda\dot\gamma}\,,\qquad
    \bar H^{\dot\nu\dot\lambda}=H^{(\dot\nu\dot\lambda)}:=h^{\ \dot\nu}_{\gamma} h^{\gamma\dot\lambda}\,.
\end{align}
In the $\eta$-sector only the second term of (\ref{e: HH_decomp}) is nontrivial. One then gets

	\begin{equation}\label{e: OOC+COO}
		\Omega \bar{*} \Omega \bar{*} C \bigg|_{\bar{H}\bar{\partial}^2} =  C \bar{*} \Omega \bar{*} \Omega \bigg|_{\bar{H}\bar{\partial}^2} = \frac{1}{8} \bar H^{\dot\mu\dot\nu} y_{1\nu} y_2^\nu \bar{\partial}_{\dot{\mu}}\bar{\partial}_{\dot{\nu}}C(0,\bar{y}; k, \bar{k})\,,
	\end{equation}
	\begin{equation}\label{e: OCO}
		\Omega \bar{*} C \bar{*} \Omega \bigg|_{\bar{H}\bar{\partial}^2} =  -\frac{1}{8} \bar H^{\dot\mu\dot\nu} y_{1\nu} y_2^\nu \bar{\partial}_{\dot{\mu}}\bar{\partial}_{\dot{\nu}}C(0,\bar{y}; -k, -\bar{k})\,.
	\end{equation}
         The role of the  auxiliary variables $y_{1,2}$ is that the action of bilinears $y^\mu t_i{}_\mu$, $p^\mu t_i{}_\mu$ and $t_1^\mu t_2{}_\mu$ replaces $y_{1,2}$ with actual variables $y^\mu$, derivatives $p^\mu$ or organizes the index contraction via $t_1^\mu t_2{}_\mu$. 	In all vertices the pre-exponent contains one of the  bilinear factors $y^\mu t_i{}_\mu$, $p^\mu t_i{}_\mu$ or $t_1^\mu t_2{}_\mu$. The action of $y^\mu t_i{}_\mu$ and $p^\mu t_i{}_\mu$ on two $\Omega$ and  single $C$ yields
	\begin{equation}
		y^\mu t_1{}_\mu \Omega \bar{*} \Omega \bar{*} C \bigg|_{\bar{H}\bar{\partial}^2} = y^\mu t_1{}_\mu C \bar{*} \Omega \bar{*} \Omega \bigg|_{\bar{H}\bar{\partial}^2} =  \frac{i}{8}\bar H^{\dot\mu\dot\nu} y_{\nu} y_2^\nu \bar{\partial}_{\dot{\mu}}\bar{\partial}_{\dot{\nu}}C(0,\bar{y};k, \bar{k})\,,
	\end{equation}
	\begin{equation}
		y^\mu t_1{}_\mu \Omega \bar{*} C \bar{*} \Omega \bigg|_{\bar{H}\bar{\partial}^2} = - \frac{i}{8} \bar H^{\dot\mu\dot\nu} y_{\nu} y_2^\nu \bar{\partial}_{\dot{\mu}}\bar{\partial}_{\dot{\nu}}C(0,\bar{y};- k, -\bar{k})\,,
	\end{equation}
        \begin{equation}
		p^\mu t_1{}_\mu \Omega \bar{*} \Omega \bar{*} C \bigg|_{\bar{H}\bar{\partial}^2} = p^\mu t_1{}_\mu C \bar{*} \Omega \bar{*} \Omega \bigg|_{\bar{H}\bar{\partial}^2} =  \frac{i}{8}\bar H^{\dot\mu\dot\nu} p_{\nu} y_2^\nu \bar{\partial}_{\dot{\mu}}\bar{\partial}_{\dot{\nu}}C(0,\bar{y};k, \bar{k})\,,
	\end{equation}
	\begin{equation}
		p^\mu t_1{}_\mu \Omega \bar{*} C \bar{*} \Omega \bigg|_{\bar{H}\bar{\partial}^2} = - \frac{i}{8} \bar H^{\dot\mu\dot\nu} p_{\nu} y_2^\nu \bar{\partial}_{\dot{\mu}}\bar{\partial}_{\dot{\nu}}C(0,\bar{y};- k, -\bar{k})\,.
	\end{equation}
	For  $y^\mu t_2{}_\mu$ and $p^\mu t_2{}_\mu$ the situation is analogous up to the exchange of $y_2$ with $y_1$ and an additional minus sign. Examining the
exponents in all  vertices, constructed in Section \ref{Mod Shift}, one observes that it is impossible to
obtain the desired form of the First On-Shell Theorem from the terms with $y^\mu t_i{}_\mu$ and $p^\mu t_i{}_\mu$ in the
 pre-exponent
  since the First On-Shell Theorem does not contain terms with $y_\mu$ or $p_\mu$ contracted with
   the frame field $h$. The combination $t_1^\mu t_2{}_\mu$ leads to the correct contraction
 of two frame fields %$h_\lambda^{\ \dot\mu}\wedge h^{\lambda\dot\nu}\,\bar\dd_{\dot\mu}\bar\dd_{\dot\nu} C$:
	\begin{equation}
		t_1^\mu t_2{}_\mu \Omega \bar{*} \Omega \bar{*} C \bigg|_{\bar{H}\bar{\partial}^2} = t_1^\mu t_2{}_\mu C \bar{*} \Omega \bar{*} \Omega \bigg|_{\bar{H}\bar{\partial}^2} = -\frac{1}{4} \bar H^{\dot\mu\dot\nu} \bar{\partial}_{\dot{\mu}}\bar{\partial}_{\dot{\nu}}C(0,\bar{y}; k, \bar{k})\,,
	\end{equation}
	\begin{equation}
		t_1^\mu t_2{}_\mu \Omega \bar{*} C \bar{*} \Omega \bigg|_{\bar{H}\bar{\partial}^2} = \frac{1}{4}\bar H^{\dot\mu\dot\nu} \bar{\partial}_{\dot{\mu}}\bar{\partial}_{\dot{\nu}}C(0,\bar{y}; -k, -\bar{k})\,.
	\end{equation}
	Plugging these expressions into the vertex components
(\ref{e: preexp_first})-(\ref{e: preexp_last}), that contain the pre-exponential factor $t_1^\mu t_2{}_\mu$, and integrating out $\tau_i$ we obtain
        \begin{multline}
            \Omega*\Omega*C*h_{n_1 t_1 + \beta_1 + (1-\beta_1)(p+t_1+t_2)}\Delta_{\alpha y + (1-\alpha+\lambda)(p+t_2)}\Delta_{\alpha y + (1-\alpha+\lambda)(p+t_1+t_2)}\gamma\bigg|_{\bar{H}\bar{\partial}^2} = \frac{\lambda (\alpha - 1 - \lambda)}{4}\bar H^{\dot\mu\dot\nu} \bar{\partial}_{\dot{\mu}}\bar{\partial}_{\dot{\nu}}\\
            \bigg[C(0,\bar{y};k, \bar{k}) +  2 \frac{\beta_1^{n+2} + (\alpha-\lambda)^{n+1}((\alpha-\lambda)(n+1)-\beta_1(n+2))}{(n+1)(n+2)(\alpha-\lambda - \beta_1)^2}y^{\mu_1}...y^{\mu_n}\bar{y}^{\dot{\mu}_1}...\bar{y}^{\dot{\mu}_m}C_{\mu_1...\mu_n, \dot{\mu}_1...\dot{\mu}_m}(k, \bar{k}) \bigg]k\,,
        \end{multline}
	\begin{multline}
		\Omega*\Omega*C*h_{n_1 t_2 + (1-\beta_1)(p + t_2)}\Delta_{\alpha y + (1-\alpha+\lambda)p}\Delta_{\alpha y + (1-\alpha+\lambda)(t_2 + p)} \gamma \bigg|_{\bar{H}\bar{\partial}^2}  = -\frac{(\alpha -\beta_1)(\alpha - 1 - \lambda)}{4} \bar H^{\dot\mu\dot\nu} \bar{\partial}_{\dot{\mu}}\bar{\partial}_{\dot{\nu}}\\
        \bigg[C(0,\bar{y};k, \bar{k}) +  2 \frac{\beta_1^{n+2} + (\alpha-\lambda)^{n+1}((\alpha-\lambda)(n+1)-\beta_1(n+2))}{(n+1)(n+2)(\alpha-\lambda - \beta_1)^2}y^{\mu_1}...y^{\mu_n}\bar{y}^{\dot{\mu}_1}...\bar{y}^{\dot{\mu}_m}C_{\mu_1...\mu_n, \dot{\mu}_1...\dot{\mu}_m}(k, \bar{k}) \bigg]k\,,
	\end{multline}
	\begin{multline}
		\Omega*\Omega*C*h_{\alpha y + (1-\alpha+\lambda)p}\Delta_{\alpha y + (1-\alpha+\lambda)(t_1+t_2+p)}\Delta_{\alpha y + (1-\alpha + \lambda)(t_2 + p)}\gamma \bigg|_{\bar{H}\bar{\partial}^2}  = \\ =  \frac{(1-\alpha+ \lambda)^2}{4} \bar H^{\dot\mu\dot\nu} \bar{\partial}_{\dot{\mu}}\bar{\partial}_{\dot{\nu}}C((\alpha-\lambda) y,\bar{y};k, \bar{k})k\,,
	\end{multline}
	\begin{multline}
		\Omega*C*\Omega*h_{\alpha y + (1-\alpha+\lambda)(t_1+t_2+p)}\Delta_{\alpha y + (1-\alpha+\lambda)(t_1+p) + 2t_2}\Delta_{\alpha y + (1-\alpha+\lambda)(t_2+p)}\gamma \bigg|_{\bar{H}\bar{\partial}^2}= \\ = \frac{((\alpha-\lambda)^2-1)}{4}
		\bar H^{\dot\mu\dot\nu} \bar{\partial}_{\dot{\mu}}\bar{\partial}_{\dot{\nu}}C((\alpha-\lambda) y,\bar{y};-k, -\bar{k})k\,,
	\end{multline}
	\begin{multline}
		\Omega*C*\Omega*h_{\alpha y + (1-\alpha+\lambda)(t_1+p) + 2t_2}\Delta_{\alpha y + (1-\alpha+\lambda)p + 2t_2}\Delta_{\alpha y+(1-\alpha+\lambda)(t_2+p)}\gamma \bigg|_{\bar{H}\bar{\partial}^2} = \\ = \frac{((\alpha-\lambda)^2-1)}{4}
		\bar H^{\dot\mu\dot\nu} \bar{\partial}_{\dot{\mu}}\bar{\partial}_{\dot{\nu}}C((\alpha-\lambda) y,\bar{y};-k, -\bar{k})k\,,
	\end{multline}
	\begin{multline}
		\Omega*C*\Omega*h_{n_1 t_1 + \beta_1 y + (1-\beta_1)(p+t_1) + 2t_2}\Delta_{\alpha y + (1-\alpha+\lambda)p + 2t_2}\Delta_{\alpha y+(1-\alpha+\lambda)(t_1+p) + 2t_2}\gamma \bigg|_{\bar{H}\bar{\partial}^2}= \\ = -\frac{(\alpha - \beta_1)(\alpha - 1 - \lambda)}{4} \bar H^{\dot\mu\dot\nu} \bar{\partial}_{\dot{\mu}}\bar{\partial}_{\dot{\nu}} \bigg[C(0,\bar{y};-k, -\bar{k}) + \\ +  2 \frac{\beta_1^{n+2} + (\alpha-\lambda)^{n+1}((\alpha-\lambda)(n+1)-\beta_1(n+2))}{(n+1)(n+2)(\alpha-\beta_1 -\lambda)^2}y^{\mu_1}...y^{\mu_n}\bar{y}^{\dot{\mu}_1}...\bar{y}^{\dot{\mu}_m} C_{\mu_1...\mu_n, \dot{\mu}_1...\dot{\mu}_m}(-k, -\bar{k}) \bigg]k\,,	
	\end{multline}
	\begin{multline}
		\Omega*C*\Omega*h_{n_1 t_1 + \beta_1 y + (1-\beta_1)(p+t_1+t_2) }\Delta_{\alpha y + (1-\alpha+\lambda)(p+t_1 + t_2)}\Delta_{\alpha y+(1-\alpha+\lambda)(t_2+p)}\gamma \bigg|_{\bar{H}\bar{\partial}^2}= \\ = \frac{\lambda (\alpha - 1 - \lambda)}{4} \bar H^{\dot\mu\dot\nu} \bar{\partial}_{\dot{\mu}}\bar{\partial}_{\dot{\nu}} \bigg[C(0,\bar{y};-k, -\bar{k}) + \\ +  2 \frac{\beta_1^{n+2} + (\alpha-\lambda)^{n+1}((\alpha-\lambda)(n+1)-\beta_1(n+2))}{(n+1)(n+2)(\alpha-\beta_1 -\lambda)^2}y^{\mu_1}...y^{\mu_n}\bar{y}^{\dot{\mu}_1}...\bar{y}^{\dot{\mu}_m} C_{\mu_1...\mu_n, \dot{\mu}_1...\dot{\mu}_m}(-k, -\bar{k}) \bigg]k\,,	
	\end{multline}
	\begin{multline}
		\Omega*C*\Omega*h_{n_2 t_2+ \beta_2 y + (1-\beta_2)(p + t_2)}\Delta_{\alpha y + (1-\alpha+\lambda)(t_2+p)}\Delta_{\alpha y + (1-\alpha+\lambda)p + 2t_2}\gamma \bigg|_{\bar{H}\bar{\partial}^2} = \\ = -\frac{(\alpha - \beta_2)(\alpha + 1-\lambda)}{4} \bar H^{\dot\mu\dot\nu} \bar{\partial}_{\dot{\mu}}\bar{\partial}_{\dot{\nu}} \bigg[C(0,\bar{y};-k, -\bar{k}) + \\ +  2 \frac{\beta_2^{n+2} + (\alpha-\lambda)^{n+1}((\alpha-\lambda)(n+1)-\beta_2(n+2))}{(n+1)(n+2)(\alpha-\beta_2 - \lambda)^2}y^{\mu_1}...y^{\mu_n}\bar{y}^{\dot{\mu}_1}...\bar{y}^{\dot{\mu}_m} C_{\mu_1...\mu_n, \dot{\mu}_1...\dot{\mu}_m}(-k, -\bar{k}) \bigg]k\,,	
	\end{multline}
		\begin{multline}
		\Omega*C*\Omega*h_{n_2 t_2+ \beta_2 y + (1-\beta_2)(p + t_1 + t_2)}\Delta_{\alpha y + (1-\alpha+\lambda)(t_1+p) +2t_2}\Delta_{\alpha y + (1-\alpha+\lambda)(p + t_1+ t_2)}\gamma \bigg|_{\bar{H}\bar{\partial}^2} = \\ = \frac{\lambda(\alpha + 1-\lambda)}{4} \bar H^{\dot\mu\dot\nu} \bar{\partial}_{\dot{\mu}}\bar{\partial}_{\dot{\nu}} \bigg[C(0,\bar{y};-k, -\bar{k}) + \\ +  2 \frac{\beta_2^{n+2} + (\alpha-\lambda)^{n+1}((\alpha-\lambda)(n+1)-\beta_2(n+2))}{(n+1)(n+2)(\alpha-\beta_2 - \lambda)^2}y^{\mu_1}...y^{\mu_n}\bar{y}^{\dot{\mu}_1}...\bar{y}^{\dot{\mu}_m} C_{\mu_1...\mu_n, \dot{\mu}_1...\dot{\mu}_m}(-k, -\bar{k}) \bigg]k\,,	
	\end{multline}
	\begin{multline}
	    C*\Omega*\Omega*h_{n_2 t_2 + \beta_2 y + (1-\beta_2)(p+t_1+t_2)}\Delta_{\alpha y + (1-\alpha+\lambda)(p+t_1+t_2)}\Delta_{\alpha y + (1-\alpha+\lambda)(p+t_1) + 2 t_2}\gamma \bigg|_{\bar{H}\bar{\partial}^2}= \\ = \frac{\lambda(\alpha+1-\lambda)}{4}
        \bar H^{\dot\mu\dot\nu} \bar{\partial}_{\dot{\mu}}\bar{\partial}_{\dot{\nu}} \bigg[C(0,\bar{y};k, \bar{k}) + \\ +  2 \frac{\beta_2^{n+2} + (\alpha-\lambda)^{n+1}((\alpha-\lambda)(n+1)-\beta_2(n+2))}{(n+1)(n+2)(\alpha-\beta_2-\lambda)^2}y^{\mu_1}...y^{\mu_n}\bar{y}^{\dot{\mu}_1}...\bar{y}^{\dot{\mu}_m} C_{\mu_1...\mu_n, \dot{\mu}_1...\dot{\mu}_m}(k, \bar{k}) \bigg]k\,,
	\end{multline}
        \begin{multline}
		C*\Omega*\Omega*h_{n_2 t_1 + \beta_2 y + (1-\beta_2)(p + t_1) + 2t_2}\Delta_{\alpha y + (1-\alpha+\lambda)(t_1+p) + 2t_2}\Delta_{\alpha y + (1-\alpha+\lambda)p +2t_1 +2t_2}\gamma \bigg|_{\bar{H}\bar{\partial}^2} = \\ = -\frac{(\alpha+1-\lambda)(\alpha-\beta_2)}{4} \bar H^{\dot\mu\dot\nu} \bar{\partial}_{\dot{\mu}}\bar{\partial}_{\dot{\nu}} \bigg[C(0,\bar{y};k, \bar{k}) + \\ +  2 \frac{\beta_2^{n+2} + (\alpha-\lambda)^{n+1}((\alpha-\lambda)(n+1)-\beta_2(n+2))}{(n+1)(n+2)(\alpha-\beta_2-\lambda)^2}y^{\mu_1}...y^{\mu_n}\bar{y}^{\dot{\mu}_1}...\bar{y}^{\dot{\mu}_m} C_{\mu_1...\mu_n, \dot{\mu}_1...\dot{\mu}_m}(k, \bar{k}) \bigg]k\,,	
	\end{multline}
	\begin{multline}
		C*\Omega*\Omega*h_{\alpha y + (1-\alpha+\lambda)p +2t_1 +2t_2}\Delta_{\alpha y + (1-\alpha+\lambda)(t_1+p) + 2t_2}\Delta_{\alpha y + (1-\alpha+\lambda)(t_1+t_2+p)}\gamma \bigg|_{\bar{H}\bar{\partial}^2} = \\ =  \frac{(1+\alpha-\lambda)^2}{4} \bar H^{\dot\mu\dot\nu} \bar{\partial}_{\dot{\mu}}\bar{\partial}_{\dot{\nu}} C((\alpha - \lambda) y,\bar{y};k, \bar{k})k \,.
	\end{multline}
	As a result,
	\begin{multline}\label{e: e+o_sum}
		\Upsilon^\eta(\Omega,\Omega,C) + \Upsilon^\eta(\Omega,C,\Omega) + \Upsilon^\eta(C,\Omega,\Omega)\bigg|_{\bar{H}\bar{\partial}^2} =  \frac{\eta}{8i} \bar H^{\dot\mu\dot\nu} \bar{\partial}_{\dot{\mu}}\bar{\partial}_{\dot{\nu}}\bigg[((\alpha-\lambda)^2 + 1)C((\alpha - \lambda) y,\bar{y};k, \bar{k}) + \\
        + ((\alpha-\lambda)^2 - 1) C((\alpha-\lambda) y,\bar{y};-k, -\bar{k}) \bigg]k - \frac{\eta}{16 i} \bar H^{\dot\mu\dot\nu} \bar{\partial}_{\dot{\mu}}\bar{\partial}_{\dot{\nu}} \Bigg[(\alpha-\beta_1-\lambda)(\alpha-1-\lambda) + \\ + (\alpha-\beta_2-\lambda)(\alpha+1-\lambda)\Bigg]C(0,\bar{y};k, \bar{k})k
		- \frac{\eta}{16 i} \bar H^{\dot\mu\dot\nu} \bar{\partial}_{\dot{\mu}}\bar{\partial}_{\dot{\nu}} \Bigg[(\alpha-\beta_1-\lambda)(\alpha-1-\lambda) + \\ + (\alpha-\beta_2-\lambda)(\alpha+1-\lambda)\Bigg]C(0,\bar{y};-k, -\bar{k})k - \\ - \frac{\eta(\alpha-1-\lambda)}{8i}\frac{\beta_1^{n+2} + (\alpha-\lambda)^{n+1}((\alpha-\lambda)(n+1)-\beta_1(n+2))}{(n+1)(n+2)(\alpha-\beta_1-\lambda)}\\ \bar H^{\dot\mu\dot\nu} \bar{\partial}_{\dot{\mu}}\bar{\partial}_{\dot{\nu}}y^{\mu_1}...y^{\mu_n}\bar{y}^{\dot{\mu}_1}...\bar{y}^{\dot{\mu}_m}
        \Bigg[C_{\mu_1...\mu_n, \dot{\mu}_1...\dot{\mu}_m}(k, \bar{k}) +  C_{\mu_1...\mu_n, \dot{\mu}_1...\dot{\mu}_m}(-k, -\bar{k}) \Bigg]k - \\ - \frac{\eta(\alpha+1-\lambda)}{8i}\frac{\beta_2^{n+2} + (\alpha-\lambda)^{n+1}((\alpha-\lambda)(n+1)-\beta_2(n+2))}{(n+1)(n+2)(\alpha-\beta_2-\lambda)}\\
        \bar H^{\dot\mu\dot\nu} \bar{\partial}_{\dot{\mu}}\bar{\partial}_{\dot{\nu}}y^{\mu_1}...y^{\mu_n}\bar{y}^{\dot{\mu}_1}...\bar{y}^{\dot{\mu}_m}\Bigg[C_{\mu_1...\mu_n, \dot{\mu}_1...\dot{\mu}_m}(k, \bar{k}) +  C_{\mu_1...\mu_n, \dot{\mu}_1...\dot{\mu}_m}(-k, -\bar{k}) \Bigg]k\,.
	\end{multline}

 Now, since the subsystems for the components of $C(Y;K|x)$ that are even and odd in $K$ are independent, we have to respect the First On-Shell Theorem for both physical (\ref{e: COST}) and topological fields (\ref{e: COST TOP}). This results in the doubling of shift parameters $\alpha^{e,o}, \beta_i^{e,o}, \lambda^{e,o}$.

       For odd components $C(Y;k,\bar{k}|x) = -C(Y;-k,-\bar{k}|x)$:
	\begin{equation}\label{K1}
		\Upsilon^\eta(\Omega,\Omega,C) + \Upsilon^\eta(\Omega,C,\Omega) + \Upsilon^\eta(C,\Omega,\Omega)\bigg|_{\bar{H}\bar{\partial}^2} = -\frac{i \eta}{4}\bar H^{\dot\mu\dot\nu} \bar{\partial}_{\dot{\mu}}\bar{\partial}_{\dot{\nu}}  C((\alpha^o -\lambda^o) y,\bar{y};K)k\,.
	\end{equation}
	The form of the First On-Shell Theorem in the physical sector is respected if $\alpha^o = \lambda^o$.
	
	For even components $C(Y;k,\bar{k}) = C(Y;-k,-\bar{k})$:
	\begin{multline}
		\Upsilon^\eta(\Omega,\Omega,C) + \Upsilon^\eta(\Omega,C,\Omega) + \Upsilon^\eta(C,\Omega,\Omega)\bigg|_{\bar{H}\bar{\partial}^2} = \frac{\eta}{4i}(\alpha^e - \lambda^e )^2 \bar H^{\dot\mu\dot\nu} \bar{\partial}_{\dot{\mu}}\bar{\partial}_{\dot{\nu}} C((\alpha^e -\lambda^e) y,\bar{y};K)k - \\
		-\frac{\eta}{8i}\bigg((\alpha^e - \lambda^e)(2\alpha^e + \beta_1^e - \beta_2^e - 2\lambda^e) + (\beta_1^e - \beta_2^e)\bigg)\bar H^{\dot\mu\dot\nu} \bar{\partial}_{\dot{\mu}}\bar{\partial}_{\dot{\nu}}\\
        C(0,\bar{y};K)k +  \frac{\eta}{4i}\bar H^{\dot\mu\dot\nu} \bar{\partial}_{\dot{\mu}}\bar{\partial}_{\dot{\nu}}y^{\mu_1}...y^{\mu_n}\bar{y}^{\dot{\mu}_1}...\bar{y}^{\dot{\mu}_m} C_{\mu_1...\mu_n, \dot{\mu}_1...\dot{\mu}_m}(k, \bar{k})\\
        \bigg[(\alpha^e-1-\lambda^e)\frac{(\beta_1^e)^{n+2} + (\alpha^e-\lambda^e)^{n+1}((\alpha^e-\lambda^e) (n+1)-\beta_1^e(n+2))}{(n+1)(n+2)(\alpha^e-\beta_1^e -\lambda^e)} + \\ + (\alpha^e +1-  \lambda^e)\frac{(\beta_2^e)^{n+2} + (\alpha^e-\lambda^e)^{n+1}((\alpha^e-\lambda^e)(n+1)-\beta^e_2(n+2))}{(n+1)(n+2)(\alpha^e-\beta^e_2-\lambda^e)}\bigg]k\,.
	\end{multline}
Since the First On-Shell Theorem for topological fields features no such terms, they must vanish. The decomposition of the field $C(Y;K)$ into power series in $Y$  yields an infinite chain of equations on the parameters $\alpha^e, \beta_i^e, \lambda^e$.
 \begin{multline}
     (\alpha^e-1-\lambda^e)\frac{(\beta^e_1)^{n+2} + (\alpha^e-\lambda^e)^{n+1}((\alpha^e-\lambda^e) (n+1)-\beta^e_1(n+2))}{(n+1)(n+2)(\alpha^e-\beta^e_1-\lambda^e)} + \\ + (\alpha^e+1-\lambda^e)\frac{(\beta^e_2)^{n+2} + (\alpha^e - \lambda^e)^{n+1}((\alpha^e-\lambda^e)(n+1)-\beta^e_2(n+2))}{(n+1)(n+2)(\alpha^e-\beta^e_2 - \lambda^e)} + (\alpha^e-\lambda^e)^{n+2}= 0\,, \forall n \in \mathds{N}\,,
 \end{multline}
that demand $\alpha^e = \lambda^e$ and $\beta_1^e = \beta_2^e$. The origin of these conditions is that in the  AdS background the dependence on $t_i$ can be at most bilinear, so that the terms with $t_1 t_2$ in the pre-exponent must have matching exponents at $t_i=0$ to respect the First On-Shell Theorem. Note that these constraints do not reduce the vertices to those given by the conventional homotopy for general $\omega$ which allow higher orders in $t_i$.

To find the possible solutions for $\beta_1^{e,o}$ and $\beta_2^{e,o}$, the terms in the $(y^\mu t_{i \mu} + p^\mu t_{i \mu})$ $h h\bar{\partial} \bar{\partial} C$ sector have to be inspected
\begin{multline}\label{e: yt+pt hhddC}
\Upsilon^\eta(\Omega,\Omega,C) + \Upsilon^\eta(\Omega,C,\Omega) + \Upsilon^\eta(C,\Omega,\Omega)\bigg|_{\bar{H}\bar{\partial}^2} = \\
	    	-\frac{\eta}{16} \bar H^{\dot\mu\dot\nu} \bar{\partial}_{\dot{\mu}}\bar{\partial}_{\dot{\nu}} \bigg[C_\rho(\bar{y};k,\bar{k})y^\rho\bigg(-\frac{\beta_1}{2}-\frac{\beta_2}{2}+\frac{\beta_1^2}{6}-\frac{\beta_2^2}{6} \bigg) - C_\rho(\bar{y};-k,-\bar{k})y^\rho\bigg(-\frac{\beta_1}{2}-\frac{\beta_2}{2}-\frac{\beta_1^2}{6}+\frac{\beta_2^2}{6} \bigg) - \\
      - \frac{1}{(n+1)(n+2)}C_{\rho_1...\rho_n}(\bar{y};k,\bar{k})y^{\rho_1}...y^{\rho_n}\bigg(-\beta_1^n(n+2-\beta_1 n) - \beta_2^n (n+2+\beta_2 n) \bigg) + \\
      + \frac{1}{(n+1)(n+2)}C_{\rho_1...\rho_n}(\bar{y};-k,-\bar{k})y^{\rho_1}...y^{\rho_n}\bigg(-\beta_1^n(n+2+\beta_1 n) - \beta_2^n (n+2-\beta_2 n) \bigg) \bigg]\,.
\end{multline}
Here no terms with parameters $\alpha=\lambda$ are present since they are
accompanied by the factors of the form
$a_\mu a^\mu=0$ with some spinors $a_\mu$.
Since the parameters contribute to the argument of the field $C(Y;K)$, to respect
 the First On-Shell Theorem one   has  to expand $C(Y;K)$ in power series that yields an infinite chain of algebraic equations on $\beta_1$ and $\beta_2$. The projection onto the odd sector  yields
\begin{multline}
    \Upsilon^\eta(\Omega,\Omega,C) + \Upsilon^\eta(\Omega,C,\Omega) + \Upsilon^\eta(C,\Omega,\Omega)\bigg|_{\bar{H}\bar{\partial}^2} = \\
    =  -\frac{\eta}{16} \bar H^{\dot\mu\dot\nu} \bar{\partial}_{\dot{\mu}}\bar{\partial}_{\dot{\nu}} \bigg[C_{\mu_1}y^{\mu_1}(\bar{y};k,\bar{k}) \biggl( \beta^o_1 + \beta^o_2 \biggr) + \frac{2}{(n+1)} C_{\mu_1 ... \mu_n}y^{\mu_1}...y^{\mu_n}(\bar{y};k,\bar{k}) \biggl( (\beta^o_1)^n + (\beta^o_2)^n \biggr)\bigg] \,,
\end{multline}
that only obeys the First On-Shell Theorem at $\beta_1^o = \beta_2^o = 0$.

The same reasoning in the even sector gives
\begin{equation}
    \frac{1}{3} C_{\mu_1}(\bar{y};k,\bar{k}) \biggl( (\beta^e_1)^2 - (\beta^e_2)^2 \biggr) + \frac{2n}{(n+1)(n+2)} C_{\mu_1 ... \mu_n}(\bar{y};k,\bar{k})  \biggl( (\beta^e_1)^{n+1} - (\beta^e_2)^{n+1} \biggr) = 0\,,
\end{equation}
implying $\beta^e_1 = \beta^e_2$.

  Analogous analysis can be applied to the $t^\mu_1t_2{}_\mu$ and $(y^\mu t_{i \mu} + p^\mu t_{i \mu})$-dependent terms in the $\bar{H}\bar{y}\bar{y}C$ sector. Due to the sign change in the products $\Omega\bar{*}\Omega\bar{*}C|_{\bar{Hyy}}$ and $C\bar{*}\Omega\bar{*}\Omega|_{\bar{Hyy}}$ compared to $\Omega\bar{*}\Omega\bar{*}C|_{\bar{H\partial^2}}$ and $C\bar{*}\Omega\bar{*}\Omega|_{\bar{H\partial^2}}$ we find a permutation of the even and  odd projections of a slightly changed versions of (\ref{e: e+o_sum}) and (\ref{e: yt+pt hhddC}). This yields $\alpha^{e,o} = \lambda^{e,o}$, $\beta_1^e = \beta_2^e = 0$ and $\beta_1^o = \beta_2^o$.

One can also check that the terms with $H_{\alpha\beta} y_1^\alpha y_2^\beta$
in all vertices (recall that $y_1$ and $y_2$ are the auxiliary variables assigned, respectively, to the first and second factors of $\Omega$ in the ordered product)
impose no restrictions on the parameters. The resulting restrictions on the parameters are $\alpha^{e,o} = \lambda^{e,o}$ in the both sectors, which means that they are otherwise  free. At the same time the $y$-shift parameters in $W_1$  are necessarily vanishing $\beta^{e,o}_1=\beta^{e,o}_2=0$.

The obtained results imply  that one can use two independent homotopy operators when resolving $S_1$ and $W_1$, provided the $y$ and $p$ shifts are equal within each homotopy procedure:
\begin{align}
    & S_1 = - \frac{\eta}{2} \Delta_{a(y+p)} \big(C * \gamma \big) + c.c. \,,\\
    & W_1 = - \frac{i}{2} \Delta_{b(y+p)} (d_x S_1 + \omega * S_1 + S_1 * \omega) + c.c. \,
\end{align}
with independent $a$ and $b$. Such a homotopy procedure generalizes  uniform shifts considered in \cite{Didenko:2018fgx}, where only the shifts with $a=b$ were considered, that preserve the form  of the conventional homotopy vertices. The case of different $a$ and $b$ is referred to as the \textit{relaxed} uniform shift.
 We have shown that the relaxed uniform shifts  produce vertices that differ from those resulting from the conventional homotopy in general HS background but still respect 
 the First On-Shell Theorem in $AdS_4$ background.

It is worth noticing that $n_i$-parameters are not present in the above considerations. This suggests that there is no interplay between the $y, p$-shifts and $\omega$-shifts, which indicates that the latter do not affect the First On-Shell Theorem at all. In the particular case of a pure $\omega$-shift ($\alpha = \lambda = \beta_i = 0$) these parameters do not
contribute even beyond the level of free HS equations in $AdS_4$, as they do not affect the HS fields, being equivalent to those resulting from the conventional ({\it i.e.} zero shift) homotopy.

	\section{Pure $\omega$-shift }\label{Omega Shift}
	
	Now we consider the effect of the pure shift by the arguments of $\omega$ on
 the full $\omega^2 C$ vertices beyond  the $AdS_4$
background. To this end we set $q^\mu = l_i^\mu  = \beta_i = \alpha = \lambda = 0$ leaving the $\omega$--shifts  with  parameters $n_i$ free.
	From pre-exponential factors (\ref{e: preexp_first}) - (\ref{e: preexp_last})  one can see that the only non-zero terms at $\alpha = \lambda =\beta_i=0$ are
	
	\begin{equation}\label{e: VOOC_ZERO_FULL}
		\Upsilon^\eta(\omega,\omega,C) = \frac{\eta}{4i} \omega*\omega*C*h_{p}\Delta_{t_1+t_2+p}\Delta_{t_2 + p}\gamma\,,
	\end{equation}
	\begin{equation}\label{e: VOCO_ZERO_FULL}
		\Upsilon^\eta(\omega,C,\omega) = \frac{\eta}{4i} \omega*C*\omega*\big[h_{t_1+t_2+p}\Delta_{t_1+p+2t_2}\Delta_{t_2+p}\gamma + h_{t_1+p+2t_2}\Delta_{p + 2t_2}\Delta_{t_2+p}\gamma\big]\,,
	\end{equation}
	\begin{equation}\label{e: VCOO_ZERO_FULL}
		\Upsilon^\eta(C,\omega,\omega) = \frac{\eta}{4i} C*\omega*\omega*h_{p +2t_1 +2t_2}\Delta_{t_1 + p + 2t_2}\Delta_{t_1+t_2+p}\gamma\,.
	\end{equation}
	This yields the equation
	\begin{multline}\label{e: ZERO_SHIFT_EQ}
		\d \omega + \omega*\omega + \omega*\omega*C*h_{p}\Delta_{t_1+t_2+p}\Delta_{t_2 + p}\gamma + \omega*C*\omega*h_{t_1+t_2+p}\Delta_{t_1+p+2t_2}\Delta_{t_2+p}\gamma  + \\ + \omega*C*\omega*
		h_{t_1+p+2t_2}\Delta_{p + 2t_2}\Delta_{t_2+p}\gamma +  C*\omega*\omega*h_{p +2t_1 +2t_2}\Delta_{t_1+p+2t_2}\Delta_{t_1+t_2+p}\gamma= 0\,.
	\end{multline}
Using (\ref{e: hdd_int}) and partial star-product (\ref{bar_star_product}) we obtain
	\begin{multline}\label{e: VOOC_ZERO}
		\Upsilon^\eta(\omega, \omega, C)=\frac{\eta}{2 i} \int_{[0,1]^{3}} d^{3} \tau \delta\left(1-\tau_{1}-\tau_{2}-\tau_{3}\right) e^{i\left(1-\tau_{3}\right) \partial_{1}^{\mu} \partial_{2 \mu}} \partial^{\nu} \omega\left(\left(1-\tau_{1}\right) y, \bar{y}; K\right) \bar * \\ \bar * \partial_{\nu} \omega\left(\tau_{2} y, \bar{y}; K\right)\bar *
		 C\left(-i \tau_{1} \partial_{1}-i\left(1-\tau_{2}\right) \partial_{2}, \bar{y} ; K\right) k\,,
	\end{multline}
	\begin{multline}	\label{e: VCOO_ZERO}
		\Upsilon^\eta(C, \omega, \omega)=\frac{\eta}{2 i} \int_{[0,1]^{3}} d^{3} \tau \delta\left(1-\tau_{1}-\tau_{2}-\tau_{3}\right) e^{i\left(1-\tau_{3}\right) \partial_{1}^{\mu} \partial_{2 \mu}} \\ C\left(i \tau_{1} \partial_{2}+i\left(1-\tau_{2}\right) \partial_{1}, \bar{y} ; K\right)
		\bar * \partial^{\nu} \omega\left(\tau_{2} y, \bar{y}; K\right) \bar * \partial_{\nu} \omega\left(-\left(1-\tau_{1}\right) y, \bar{y}; K\right) k\,,
	\end{multline}
	\begin{multline}\label{e: VOCO_ZERO}
		\Upsilon^\eta(\omega, C, \omega)=\frac{\eta}{2 i} \int_{[0,1]^{3}} d^{3} \tau \delta\left(1-\tau_{1}-\tau_{2}-\tau_{3}\right) e^{i\left(1-\tau_{3}\right) \partial_{1}^{\mu} \partial_{2 \mu}} \partial^{\nu} \omega\left(\tau_{1} y, \bar{y}; K\right) \bar * \\
		\bar * C\left(i\left(1-\tau_{2}\right) \partial_{2}-i\left(1-\tau_{1}\right) \partial_{1}, \bar{y} ; K\right) \bar *  \partial_{\nu} \omega\left(-\left(1-\tau_{2}\right) y, \bar{y}; K\right) k + \\ + \frac{\eta}{2 i} \int_{[0,1]^{3}} d^{3} \tau
		\delta\left(1-\tau_{1}-\tau_{2}-\tau_{3}\right) e^{-i \tau_{2} \partial_{1}^{\mu} \partial_{2 \mu}} \partial^{\nu} \omega\left(\left(1-\tau_{1}\right) y, \bar{y}; K\right)  \bar * \\ \bar	* C\left(-i \tau_{1} \partial_{1}+i \tau_{3} \partial_{2}, \bar{y} ; K\right)
		\bar * \partial_{\nu} \omega\left(-\left(1-\tau_{3}\right) y, \bar{y}; K\right) k\,.
	\end{multline}
	Remarkably,  $n_{1,2}$ do not contribute to the vertices (\ref{e: VOOC_ZERO})-(\ref{e: VOCO_ZERO}),
 which  coincide with those resulting from the conventional homotopy procedure with
 zero shift parameters
 \cite{Didenko:2018fgx}.
	
	The same result can be obtained in a simpler way using the property of the $\Delta_Q$ and $h_Q$
(\ref{e: hdd}) presented in Section \ref{Shifted homotopies}. Moreover, the absence of restrictions
on the parameters $n_i$ can already be established at the level  of the field $W_1^\eta$.
	For instance, consider  the field $W_1^{(1)}$  with the parameter $n_1$
		\begin{equation}
		W^{\eta(1)}_1 = \frac{\eta}{4i} \omega*C*\Delta_{(n_1 + 1) t + p}\big(\Delta_{p} - \Delta_{(t+p)}\big)\gamma\,.
	\end{equation}
	Using (\ref{e: hdd}), one gets
	\begin{equation}
		W^{\eta(1)}_1 = \frac{\eta}{4i} \omega*C*\big(h_{(n_1 + 1) t + p}\Delta_{p} \Delta_{(t+p)} - \Delta_{p} \Delta_{(t+p)}\big)\gamma\,.
	\end{equation}
 Inspecting the seemingly $n_1$-dependent
 first term, we find that it vanishes by virtue of (\ref{hdd=0}) which proves
 independence of  $W^{\eta(1)}_1$ of $n_1$. Analogously, $W_1^{\eta(2)}$ is $n_2$-independent. Thus, for any
 $n_i$, the field $W_1^\eta$ is the same as in the case of the conventional homotopy, i.e. at $n_{1,2}=0$, and
 the form of the First On-Shell Theorem is intact. The output of this analysis is  that, being equivalent to the conventional homotopy,  pure
 $\omega$-shifts do not affect higher-order corrections to the fields and  non-linear HS equations.

	\section{Conclusion}\label{Concl}
	
	In this paper we have analysed an extension of the homotopy procedure elaborated in
\cite{Didenko:2018fgx} to the homotopy operators with the shift parameters acting on the
arguments of the one-form HS gauge fields $\omega$, the arguments of the zero-form HS fields $C$ and proportional to spinor variables $Y^A$.
We have found general restrictions on the shift parameters
that respect  the canonical form of the free unfolded HS equations known as First On-Shell theorem \cite{Vasiliev:1989},
	which is necessary to preserve the interpretation of zero-forms $C$ as derivatives of the
 HS gauge fields. To put it differently, this condition is demanded to provide locality
  of the unfolded HS equations at the free field level. The conditions that respect  canonical form
  of the First On-Shell theorem are shown to leave six free parameters ($n_i^{e,o}$, $\alpha^{e,o}$), four of which ($n_i^{e,o}$) are associated with the shifts of the arguments of the one-form $\omega$, and  the other two  ($\alpha^{e,o}$) of the  $(p+y)$-shift.

Thus, in the perturbative analysis, one can use different homotopy operators $\Delta_{a(y+p)}$ and $\Delta_{n_i t + b(y+p)}$ to resolve for $S_1$ and $W_1$, respectively, still preserving the form of the  First On-Shell Theorem. In the particular case of $y$ and $p$ shifts, this results generalize the uniform shifts of \cite{Didenko:2018fgx} with $a=b$. 
Relaxing this condition to \textit{relaxed uniform shifts} with independent $a$ and $b$
we have shown that the relaxed uniform shifts  produce (ultralocal) vertices that differ from those obtained by the conventional homotopy in general HS background but still respect
 the First On-Shell Theorem in $AdS_4$ background.

  In the particular case of a pure $\omega$-shift, surprisingly enough, not only the form of free HS field equations in $AdS_4$  is not
   affected by the $\omega$-shift parameters, but also all vertices $\Upsilon^\eta(\omega,\omega, C)$,
   $\Upsilon^\eta(\omega, C,\omega)$ and $\Upsilon^\eta(C,\omega,\omega)$ remain intact. Moreover,
   by virtue of identities (\ref{hdd=0}) originally obtained in
   \cite{Didenko:2018fgx} this is shown to be a consequence of the fact that
   the first-order corrections to the one-form fields $W_1^\eta(Z;Y|x)$ turn out to be independent of
   the $\omega$-shift parameters.
	
\section*{Acknowledgement}

We are grateful to Olga Gelfond, Anatoly Korybut for  useful comments and the referee
for a stimulating question.
This research was
supported by the Russian Science Foundation grant 18-12-00507.

	% Список литературы

\end{document}